\documentclass[useAMS,usenatbib, usegraphicx]{mn2e}
\usepackage{aas_macros}
\usepackage[dvips]{graphicx} 
\usepackage{txfonts} 

\title[Properties of massive compact galaxies in the Local Universe]{Young ages and other 
intriguing properties of massive compact galaxies in the Local Universe}
\author[A. Ferr\'e-Mateu et al.]{A. Ferr\'e-Mateu$^{1,2}$\thanks{E-mail:
aferre@iac.es (AFM)}, A. Vazdekis$^{1,2}$, I. Trujillo$^{1,2}$, P. S\'anchez-Bl\'azquez$^{3}$, E. Ricciardelli$^{4}$, 
\newauthor
I.G. de la Rosa$^{1,2}$ \\
$^{1}$Insituto de Astrof\'isica de Canarias, V\'ia L\'actea s/n, La Laguna, 38200, Spain\\
$^{2}$Departamento de Astrof\'isica, Universidad de La Laguna, Spain\\
$^{3}$Departamento de F\'isica Te\'orica, Universidad Aut\'onoma de Madrid, 
Spain\\
$^{4}$Departament d'Astronomia i Astrof\'isica, Universitat de Val\`{e}ncia}
\begin{document}

\date{Accepted 2012 March 12. Received 2012 February 24; in original form 2011 December 19}

\pagerange{\pageref{firstpage}--\pageref{lastpage}} \pubyear{2012}

\maketitle

\label{firstpage}

\begin{abstract}
{We characterize the kinematics, morphology, stellar populations and star formation histories of a sample of massive compact galaxies in the nearby Universe, which might provide a closer look to the nature of their high redshift (z\,$\ga$\,1.0) massive counterparts. We find that nearby compact massive objects show elongated morphologies and are fast rotators. New high-quality long-slit spectra show that they have young mean luminosity-weighted ages ($\la$\,2\,Gyr) and solar metallicities or above ([Z/H]\,$\ga$\,0.0). No significant stellar population gradients are found. The analysis of their star formation histories suggests that these objects have experienced recently enormous bursts which, in some cases, represent unprecedented large fractions of their total stellar mass. These galaxies seem to be truly unique, as they do not follow the characteristic kinematical and stellar population patterns of present-day massive ellipticals, spirals or even dwarfs.}
\end{abstract}

\begin{keywords}
galaxies: abundances --
		galaxies: evolution --
		galaxies: formation --
		galaxies: kinematics and dynamics-- 
		galaxies: stellar content
\end{keywords}

\section{Introduction}
During the last years, many studies have found that the most massive ($M_{*}\,\ga 10^{11}\,M_{\sun}$) galaxies were more compact in the past (a factor of $\sim$\,4 at z\,$>$\,1.5) than their equally massive local counterparts (\citealt{Daddi2005}, \citealt{Trujillo2006}, \citealt{Longhetti2007}, \citealt{Trujillo2007}, \citealt{Toft2007}, \citealt{Zirm2007}, \citealt{Cimatti2008}, \citealt{Buitrago2008} and others). High-z massive compact objects are the center of many scientific debates as there is not a consensus in the community about some of their properties, like their ages or velocity dispersions (\citealt{Cenarro2009}, \citealt{vanDokkum2009}, \citealt{Cappellari2009}, \citealt{Mancini2010} and others).
Thus a detailed kinematical and stellar population analysis of these objects is crucial to understand their nature. Furthermore, it is mandatory to address the question of how these high-z massive compact galaxies have evolved into the present-day massive population. There are several scenarios proposed to explain this evolution, the most popular one suggests that ``minor mergers'' are the dominant mechanism for the size and stellar mass growth (\citealt{Khochfar2006}, \citealt{Naab2006}, \citealt{Hopkins2009}, \citealt{Wuyts2010}, \citealt{Trujillo2011}, \citealt{Oser2012}). This scenario predicts that we should find few massive relic compact galaxies in the local universe having old stellar populations (\citealt{Hopkins2009}). 

Indeed, \citet{Trujillo2009} (T09 hereafter), found a sample of 48 of these superdense massive compact galaxies in the SDSS DR6 spectroscopic survey at z$<$ 0.2 (see also \citealt{Taylor2010}). An analysis of the spectra for 29 of these objects showed that they had large velocity dispersions, small effective radii and young stellar populations. This last result is in contradiction with the idea that these objects are the relics from the high-z Universe. Unfortunately the modest quality of the SDSS spectra did not allow them to perform a more detailed study of their stellar populations (e.g, gradients, star formation histories). T09 showed the need to fully characterize these local compact galaxies, not only to understand their formation, but also to understand the formation and evolution of their high-z compact analogues.\\ 

Aimed at characterizing these nearby objects, we have obtained new, high-quality spectra for seven of these compact galaxies. They are representative of the T09 sample, as they cover a range in stellar masses, magnitudes and redshifts, so their inferred properties can be extended to the majority of the nearby massive compact objects. We performed a detailed morphological, kinematical and stellar populations analysis. We derived mean SSP-equivalent ages, metallicities and abundance ratios from the absorption lines measurements and also studied their Star Formation Histories (SFH) using a full spectrum-fitting approach. In addition, we obtained high sub-kpc quality imaging with adaptive optics at GEMINI-N for some of our candidates, which confirmed their compactness and showed that their stellar surface brightness profiles are like the ones of massive compact galaxies at z\,$\sim$\,2 \citep{Trujillo2012}.\\
The layout of this paper is the following. In Sec. 2 we introduce the data and the reduction process. Section 3 describes the derived morphologies and the stellar kinematics, while Sec. 4 focuses in the stellar population analysis: line strengths (Sec. 4.1), the analysis of the gradients (Sec. 4.2) and the star formation histories (Sec. 4.3). We finally discuss both the dynamical and stellar masses of these galaxies in Sec. 5. We have adopted the cosmology of $\Omega_{m}$=\,0.3, $\Omega_{\Lambda}$=\,0.7 and H$_{0}$=\,70\,km\,s$^{-1}$ Mpc$^{-1}$.

\section[]{The Data}
The galaxies analyzed in this article were taken from the sample of 29 local bonafide massive compact galaxies from T09. Galaxies in that sample were chosen from the NYU Value-Added Galaxy Catalog (\citealt{Blanton2005}, \citealt{Blanton2007}, B07 hereafter) with 0\,$< z <$\,0.2, M$_{*}>$\,9.2\,$\times$\,10$^{10}$M$_{\sun}$ and effective radius $<$\,1.5\,kpc, adopting a Chabrier (2003) initial mass function (see T09 for more information about the sample selection). 
Note that the stellar mass, size and redshift were the only parameters considered for the sample selection in T09 and that we did not apply any further cut based on morphology, colour, or star formation rate. We obtained high-quality long-slit spectra for seven massive compact galaxies in the 4.2m William Herschel Telescope during 4 nights (22-23 July 2009 (run A) and 3-4 June 2010 (run B)). We used the blue arm of the ISIS spectrograph with the grating 600B, which gives a wavelength coverage from 3800 to 5300\,$\rm\AA{}$ at resolution 1.74\,$\rm\AA{}$ for the first run (slit-width of 1$"$) and 2.64\,$\rm\AA{}$ for the second one (slit-width of 1.5$"$), with a typical seeing of 0.7$\arcsec$. We changed the setup on the second run to improve the quality of the spectra by slightly opening the slit. Several exposures of 30 minutes were taken (see Table 1) for each galaxy depending mainly on their redshift, with the slit positioned along the major axis. Spectrophotometric standards and stars from the MILES library (\citealt{Sanchez-Blazquez2006a}, \citealt{Cenarro2007}) were observed each night for relative flux calibration. We performed a standard data reduction (bias substraction, flat-fielding, cosmic-ray removal, C- and S-distortion correction, wavelength calibration, sky substraction and flux calibration) using {\tt REDUCEME} \citep{Cardiel1999}, an optimized reduction package for long-slit spectra that gives the error that is propagated as the data is handled.

\begin{table*}
\caption{Galaxy properties}           
\label{table:1}     
\centering                      
\begin{tabular}{c|c c c c c c c c c c}   
\hline\hline      
ID NYU & R.A.(J200) & Decl.(J200) & R$_{e}$(kpc)&S\'ersic index&b/a&$\widetilde{\chi^{2}}$& z & \#exp & S/N & $\sigma$($\rm{km\,s^{-1}}$) \\    
\hline  
   54829 & 232.57104 & -0.4885095 & 1.12 & 4.60 & 0.90 & 0.721 & 0.085 & 3 & 41 &  136.8 $\pm$\,3.6\\      
   321479 & 320.21978 & 11.120310 & 1.20& 5.80 & 0.51 & 1.175 & 0.128 & 4 & 62 &  221.2 $\pm$\,5.1\\
   685469 & 335.41803 & 13.987279 & 1.48& 3.03 & 0.45 & 0.599 & 0.149 & 5 & 70 &  203.9 $\pm$\,9.9 \\
   796740 & 221.90155 & 43.496021 & 1.24& 2.40 & 0.35 & 0.566 & 0.182 & 4 & 26 &  203.3 $\pm$\,7.6 \\
   890167 & 234.89197 & 44.297863 & 0.83& 3.72 & 0.63 & 0.547 & 0.143 & 7 & 32 &  233.1 $\pm$\,11.1 \\
   896687 & 218.94667 & 54.591381 & 1.63& 5.44 & 0.95 & 0.419 & 0.130 & 4 & 50 &  222.9 $\pm$\,5.3 \\
   2434587 & 169.24737 & 17.154811 & 1.13& 5.45 & 0.40 & 0.838 & 0.172 & 7 & 46 & 205.6 $\pm$\,3.1\\
\hline                                  
\end{tabular}\\
{This table shows the main properties of the galaxies explored in this study. See T09 for other properties. The values of $\widetilde{\chi^{2}}$ is the reduced one from {\tt GALFIT}. }
\end{table*}                     
  
\section{Morphologies and Stellar Kinematics}
\subsection{Morphologies}
We used the 2-D fitting code {\tt GALFIT} (\citealt{Peng2002}, \citealt{Peng2010}) to obtain the effective radius, the S\'{e}rsic index and the axis ratio (see Table \ref{table:1}). The program makes a convolution of the S\'ersic model with the PSF of the images, determining the best fit when comparing the convolved model with the galaxy surface brightness distribution that minimizes the $\chi^{2}$ of the fit. The S\'ersic index $n$ is a measure of the shape of the surface brightness profile, giving an idea of the morphology of the object. In the nearby Universe, galaxies with $n\,<$2.5 are normally disk-like objects while $n\,>$2.5 is indicative of spheroidals \citep{Ravindranath2002}.
We have scaled the effective radius to the physical scale in kpc relative to their redshift and then circularized it:
\begin{equation}
 R_{e} = r_{e} \sqrt{b/a},
\end{equation}
where R$_{e}$ is the circularized effective radii and r$_{e}$ and \textit{b/a} the effective radii along the semi-major axis and the axial ratio derived from {\tt GALFIT}. 
Figure 1 shows, for illustration, the galaxy images, together with the model from {\tt GALFIT} and the corresponding residual for two of the galaxies (54829 and 321479). The rest of galaxies can be found in Appendix A. 
The best fit (lowest $\widetilde{\chi^{2}}$) was visually inspected, to check its quality. The majority of the galaxies in our sample are elongated (\textit{b/a}$\la$\,0.6), showing disk-like shapes. However, two of them (galaxies 54829 and 896687) are practically circular. In contrast, S\'{e}rsic indices are mainly all above 3, suggesting that they are spheroid-like objects.  With the resolution ($\sim$\,1\,arcsec) and depth of the SDSS data we are not able to trace any evidence of merging activity, although some residuals are found in the centers.

\begin{figure*}
\includegraphics[scale=0.8]{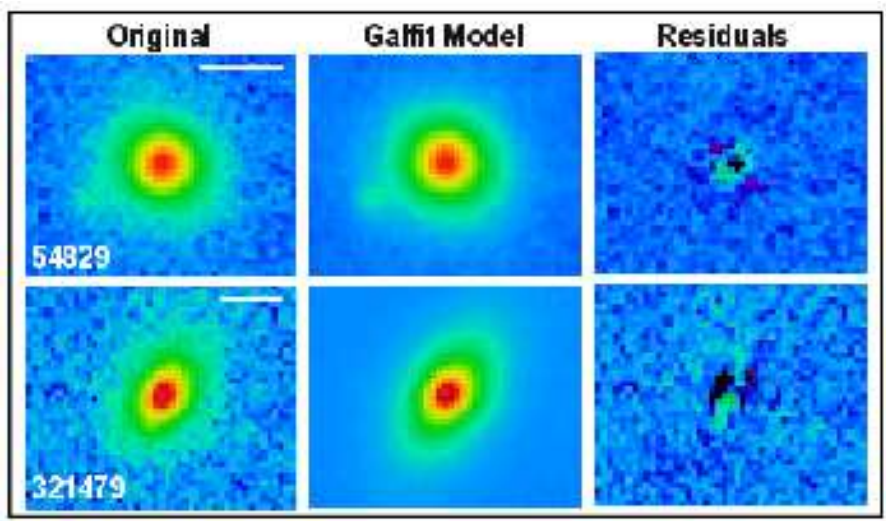}
\label{Fig.1}
\caption{Thumbnails for the compact galaxy 54829 (upper panels) and 321479 (lower panels). The first panel is the galaxy image from SDSS, the second one the fitted model from {\tt GALFIT} and finally, the residuals from the fit are shown in the third panel. The white line on the 
upper right corner of first panels is the equivalent size of 10\,kpc at the redshift of the object. The rest of the galaxies can be found in Appendix A.}
 \end{figure*} 

\subsection{Stellar Kinematics}
For the kinematical study we used the Penalized Pixel Fitting method ({\tt PPxF}; \citealt{Cappellari2004}) with the model templates from \citealt{Vazdekis2010} (V10 hereafter). To measure the radial velocities we binned in the radial direction the spectra to ensure a minimum S/N($\rm\AA{}$)=10, while we binned to achieve S/N($\rm\AA{}$)\,$\sim$\,20 for the velocity dispersion. This means that we summed at least two pixels for each bin, to take into account the effect of the seeing. In those cases where the outermost radii did not reach the required S/N we omitted those points for the analysis.\\
We find that five of our galaxies are rapidly rotating (see Fig. 2a). The galaxies 890167, 321479 and 796740 show both the largest radial velocities and higher velocity dispersions ($\sigma$). For the remaining four galaxies we cannot find a significant rotation and for some of them (54829, 685469 and 2434587),  we do not reach the flat part of the rotation curve and, therefore, we cannot calculate V$_{max}$. Considering that our galaxies could have a significant disk component due to their low \textit{b/a}, a correction for the inclination effect is explored based on the inclination derived from the axial ratio. Nevertheless, we do not have enough information to select which galaxies are face- or- edge on. Therefore, the applied correction gives us an upper limit to the rotation in case a desk exists. As shown in Fig. 2b, the maximum velocities derived from the corrected curves, are larger than those derived from the uncorrected ones. Not reaching the V$_{max}$ for some of them prevented us to conclude whether these objects are also supported by rotation. Overall, we find high radial velocities (V$_{r}$ as high as 200\,$\rm{km\,s^{-1}}$) and high velocity dispersions ($\sim$\,200\,$\rm{km\,s^{-1}}$) for most of our galaxies. The high $\sigma$ is in good agreement with the values found in T09 and in \citet{Valentinuzzi2010} for a sample of massive, similarly sized compact galaxies in local clusters (z$\sim$\,0.05).\\
Figure 2d shows the position of our objects in the anisotropy diagram (\citealt{Binney2005}, \citealt{Cappellari2007}, and references therein). This diagram relates the ratio between the ordered and random motion (V/$\sigma$) to the galaxy observed flattening ($\varepsilon$). The observed V/$\sigma$ value is obtained from the central velocity dispersion (equivalent to approximately 1\,R$_{e}$) and V$_{max}$. For those galaxies for which V$_{max}$ is not reached, we use the highest radial velocity obtained (V$_{high}$), which represents a lower limit. A dichotomy for the elliptical family has been suggested: from one side, the most massive galaxies, which are slow rotators, metal-rich, with a flat central luminosity profile, showing evidence of triaxiality; and from the other side, the less massive ones, being fast rotators, metal-poor, with a clumpy luminosity profiles, maybe containing disks and being axisymmetric (e.g., \citealt{Davies1983}, \citealt{Bender1988}, \citealt{Kormendy1996}). The two lines in Fig. 2d correspond to models for edge-on oblate galaxies with different anisotropies: the upper dotted line corresponds to isotropic models $\delta$= 0, the lower dashed one to the linear relation $\delta$= 0.7$\varepsilon_{intrinsic}$, which approximately traces the lower envelope described by the location of the observed fast-rotating galaxies on the (V/$\sigma$) diagram (see \citealt{Cappellari2007} for a full discussion of these relations). Fast rotators would lie, in this diagram, to the left of the dashed line. As it can be seen, the majority of the massive compact galaxies, without correcting them from inclination effects (big symbols) are fast rotators, except galaxies 685469 and 2434587 that fall in the slow rotators regime. However, if an inclination correction is applied, the latter galaxies practically reach the separating line. Moreover, as the V$_{max}$ was not reached for some of them, their position should be considered as a lower limit, meaning that the majority of our galaxies are fast rotators.\\

\begin{figure*}
 \centering
\includegraphics[scale=1.0]{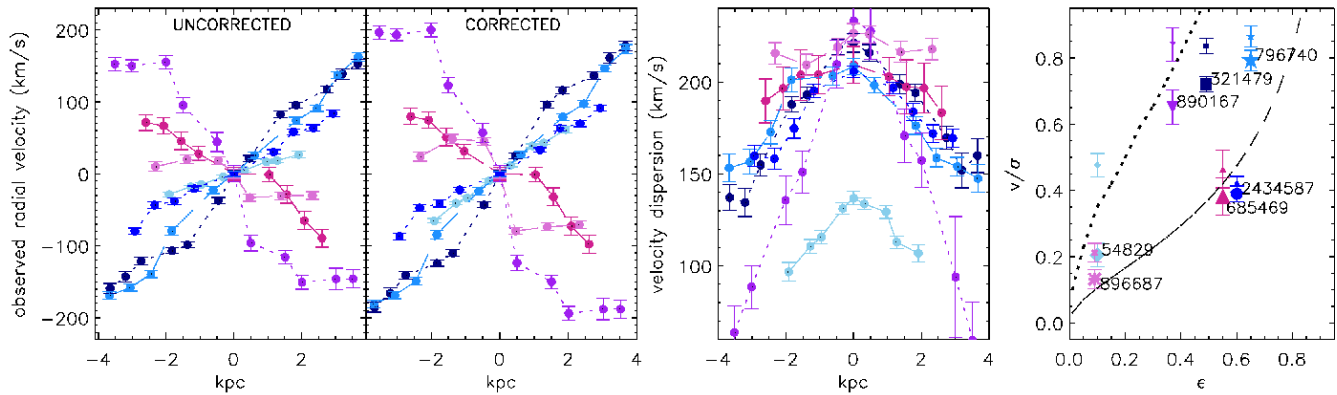}
\label{Fig.2}
\caption{Radial velocities derived with {\tt PPxF}: first panel shows the observed radial velocities, while in the second one, radial velocities corrected from inclination effects are plotted.  It can be seen that galaxies 896687 and 54829, which do not show strong rotation in the first panel, when corrected, they show slightly stronger rotations; 
\textit{b)} Velocity dispersions from {\tt PPxF}; \textit{c)} The anisotropy diagram (V/$\sigma$, $\varepsilon$). Big symbols correspond to the objects without correcting from inclination effects, while the small symbols are the same galaxies but correcting the inclination effect, in case they should be corrected. The dotted black line corresponds to the location of models for oblate edge-on isotropic ($\delta$= 0) galaxies, the dashed one to the linear relation $\delta$=\,0.7$\varepsilon_{intr}$. Relations from Cappellari et al. (2007). For those galaxies in which we do not reach the flat part of the rotation curve, these values should be considered lower limits.}
\end{figure*}

\section{Stellar Populations Analysis}
\subsection{Central Properties}
We performed a stellar population analysis using the V10 models and the newly-defined LIS-8.4\AA\, system of indices introduced in that paper. This system is characterized for having a constant resolution (FWHM=8.4\AA) across the whole wavelength range and a flux-calibrated response curve. All our line-strength measurements, including the Lick indices (\citealt{Worthey1994}, \citealt{Worthey1997}, \citealt{Trager1998}), are performed on this system. We also used the newly-defined metallicity-insensitive index H$\beta_{o}$ \citep{Cervantes2009}. This index, combined with a metallicity sensitive index, as Mgb or Fe4383, gives a very orthogonal model grid, given its reduced sensitivity to metallicity when compared to the classical H$\beta$ Lick index. The stellar population parameters were derived with {\tt rmodel} \citep{Cardiel2003}, a program that interpolates in the SSP model grids. We typically used the bivariated fit, although a careful visual inspection was also carried to check the results.\\ 
For the central stellar population analysis we summed up the spectra within $\sim$\,1\,$R_{e}$. Before summation, the galaxy spectra were corrected pixel by pixel for the radial velocity and then all pixels were broadened to the central velocity dispersion. For all galaxies, we corrected from emission line effects using {\tt GANDALF} \citep{Sarzi2006} with the V10 model templates, although we only found nebular emission in one galaxy, 685469. Figure 3 shows the index H$\beta_{o}$ \textit{versus} the combined metallicity indicator [MgFe50], that is insensitive to $\alpha$-enhancement \citep{Kuntschner2006}. Our compact galaxies have young mean SSP-equivalent ages ($\la$\,2\,Gyr). These results are in good agreement with the analysis performed in T09 but in disagreement with \citet{Valentinuzzi2010}, who found old stellar populations for their local compact massive galaxies. Overall, we find mean metallicity values above solar for our compact galaxies, though there is a scatter in the metallicity inferred in the panels of Fig.4. The galaxy 685469 is not shown in these figures. Instead, its age and metallicity were derived from the full spectrum fitting with one SSP from {\tt Ulyss} (see Sect 4.), due to the strong emission lines affecting the H$\beta_{o}$ and Fe5015 features. This galaxy is very different from the rest, being the oldest one with 2.65\,Gyr and with a total metallicity [Z/H]=\,$-$0.55. For comparison, we have overplotted some control ellipticals with similar velocity dispersions from \citet{Sanchez-Blazquez2006a} (NGC2329, NGC3379, NGC4621 and NGC5812; green crosses) and some spirals observed for the SAURON project in \citet{Falcon-Barroso2011} (NGC4235, NGC5689 and NGC6501; yellow crosses). Both control samples are in the region of the grid corresponding to old ages. Thus, our objects are much younger than both ellipticals and spirals of the local universe. This result is also shown in Figure 5, where we present the inferred ages and metallicities from our index-index plots. Our compact galaxies (purple) show much younger SSP-equivalent ages and higher total metallicities than the elliptical (green) and spiral (yellow) control samples.\\
The metallicity values derived from the different panels can be used to estimate, on a relative scale, the abundance ratio pattern of these objects. Note that the values obtained this way ([Z$_{x}$/Z$_{Fe}$]) are a good proxy for the abundance ratios (\citealt{Yamada2006}, V10). The CN was only measured for four out of the seven galaxies due to the limited spectral coverage. For these four galaxies, we find that [Z$_{CN}$/Z$_{Fe}$]$\leq$0, which is at odds with the typical [CN/Fe]$\geq$\,0 found in giant ellipticals of the same mass (see Fig.\,4). Note, however, that the [CN/Fe] shows a dependence on the environment (\citealt{Carretero2004}, \citealt{Sanchez-Blazquez2003}), with lower CN$_{2}$ for denser environments, but our objects do not belong to any known cluster or group. The [Z$_{Ca}$/Z$_{Fe}$] is also underabundant for most of our galaxies. This result is the same that the one found in \citet{Vazdekis1997} and \citet{Cenarro2004} for massive ellipticals. This behavior is not yet clear, as the Ca, which is an $\alpha$-element, is supposed to track Mg. For the [Z$_{C}$/Z$_{Fe}$] and [Z$_{Mg}$/Z$_{Fe}$], the behavior is more complex.  Visual inspection suggest that, discarding the three youngest objects that fall in the fully-degenerated part of the grid, the other galaxies do not show a significant departure from scaled-solar rate. However, the degeneracy of the models in these age regime prevents us to quantitatively determine the abundance ratio for these elements. We can conclude that the abundance ratio estimates for our compact objects, do not resemble those seen in ellipticals of similar velocity dispersion, except for the [Ca/Fe] abundance ratio. 

\begin{figure}
 \includegraphics[scale=0.7]{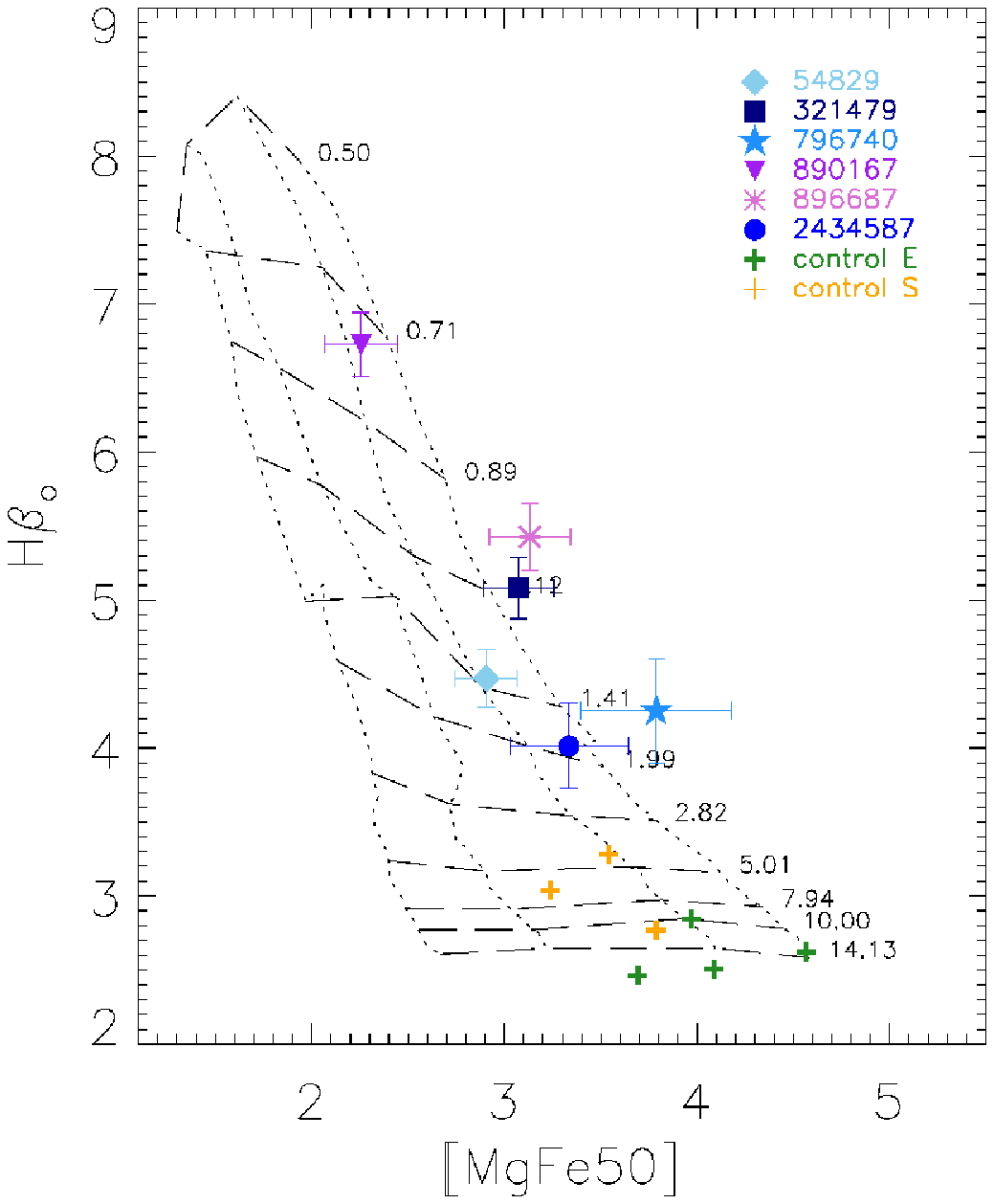}
\label{Fig.3}
\caption{The age-sensitive indicator H$\beta{_o}$. is plotted vs the composite index [MgFe50], which is a good proxy for the total metallicity. All indices are in the LIS-8.4\AA\, flux-calibrated system and the SSP model grids of V10 are plotted. Age (in Gyr) increases from top to bottom as indicated in the labels, and metallicity from left to right ([Z/H]= -0.71, -0.40, 0.00, +0.22). Symbols are like in Fig. 2 and green crosses correspond to the control ellipticals from S\'anchez-Bl\'azquez et al. (2006, paper II), that have velocity dispersions similar to those from our objects. Also overplotted as yellow crosses, the control spirals with similar velocity dispersions from the Sauron Project \citep{Falcon-Barroso2011}. Note that all the compact objects show mean SSP-equivalent ages smaller than 2\,Gyr and they all show total metallicity above solar.}
\end{figure}

\begin{figure*}
\centering 
\includegraphics{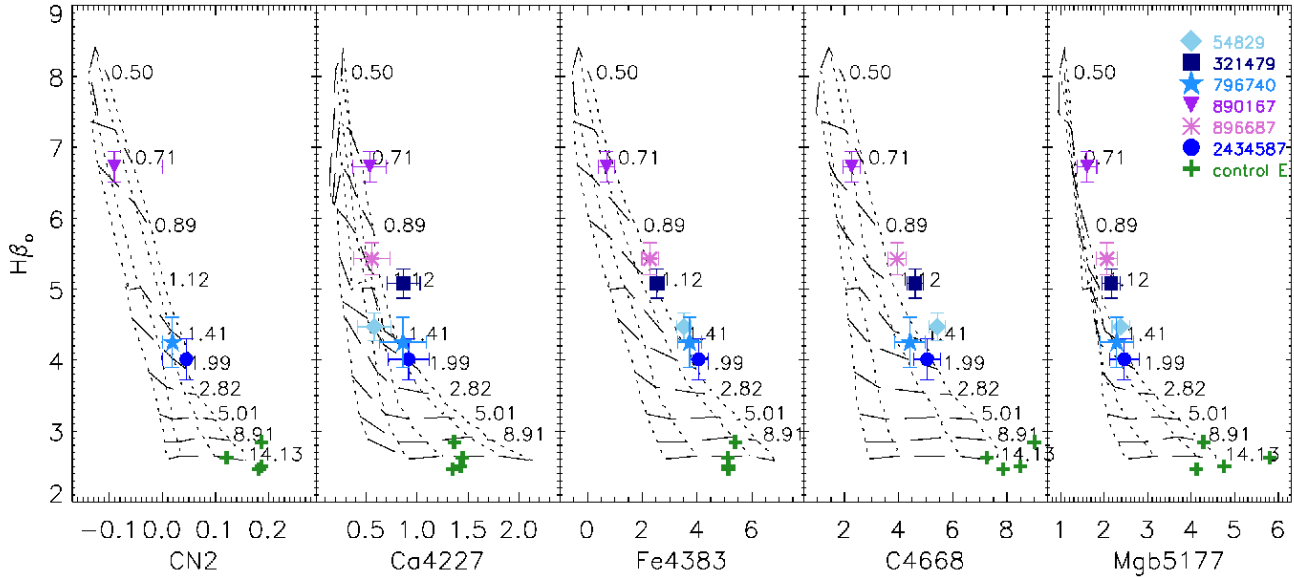}
\label{Fig.4}
\caption{The age-sensitive indicator H$\beta{_o}$ is plotted against various metallicity indices in the LIS8.4$\rm\AA{}$ flux-calibrated system. The SSP model grids of V10 are plotted. Age (in Gyr) increases from top to bottom as indicated in the labels, and metallicity from left to right ([Z/H]=-0.71, -0.40, 0.00, +0.22, except for the Ca4227 grid, where we also plot the models with [Z/H]= -1.01). Symbols correspond to Figures 2 and 3.}
\end{figure*}

\begin{table*}
\caption{Luminosity-weighted Ages and Metallicities}           
\label{table:2}     
\centering                      
\begin{tabular}{c|c c c c c c c c c c c c }   
\hline\hline      
ID NYU  & age(Gyr) & [Z/H] & age(Gyr) & [Z/H]& age(Gyr) &  [Z/H]& age(Gyr) & [Z/H] & age(Gyr) & [Z/H]\\
 & \multicolumn{2}{c}{[MgFe50]-H$\beta_{o}$} & \multicolumn{2}{c}{Fe4383-H$\beta_{o}$} & \multicolumn{2}{c}{Mgb-H$\beta_{o}$} & \multicolumn{2}{c}{C4668-H$\beta_{o}$}& \multicolumn{2}{c}{Full Spectral Fitting}\\
\hline             
   54829  & 1.35$^{+0.28}_{-0.05}$ & 0.00$^{+0.18}_{-0.30}$& 1.33$^{+0.10}_{-0.16}$ & 0.30$^{+0.08}_{-0.07}$ & 1.30$^{+0.05}_{-0.04}$ & 0.35$^{+0.05}_{-0.05}$ & 1.30$^{+0.06}_{-0.13}$ & 0.40$^{+0.10}_{-0.10}$ & 1.27$\pm$0.018 & 0.17$\pm$0.004\\      
   321479 & 1.12$^{+0.08}_{-0.06}$ & 0.30$^{+0.15}_{-0.21}$ & 1.12$^{+0.30}_{-0.12}$ & 0.22$^{+0.16}_{-0.15}$ & 1.05$^{+0.08}_{-0.05}$ & 0.50$^{+0.07}_{-0.07}$ & 1.11$^{+0.08}_{-0.06}$ & 0.35$^{+0.10}_{-0.10}$& 1.04$\pm$0.005 & 0.22$\pm$0.001\\
   796740 & 1.41$^{+0.17}_{-0.04}$ & 0.45$^{+0.12}_{-0.11}$ & 1.40$^{+0.28}_{-0.08}$ & 0.30$^{+0.18}_{-0.20}$ & 1.48$^{+0.21}_{-0.18}$ & 0.20$^{+0.05}_{-0.05}$ & 1.50$^{+0.27}_{-0.21}$ & 0.10$^{+0.10}_{-0.10}$ & 1.40$\pm$0.008 & 0.16$\pm$0.009\\
   890167 & 0.83$^{+0.07}_{-0.03}$ & 0.05$^{+0.00}_{-0.20}$ & 0.80$^{+0.08}_{-0.05}$ & 0.00$^{+0.20}_{-0.21}$ &  -  &   - & 0.80$^{+0.08}_{-0.10}$  & 0.00$^{+0.12}_{-0.25}$ & 0.65$\pm$0.006 & 0.06$\pm$0.013\\
   896687 & 1.05$^{+0.08}_{-0.06}$ & 0.40$^{+0.10}_{-0.13}$ & 0.98$^{+0.08}_{-0.06}$ & 0.25$^{+0.10}_{-0.20}$ & 0.90$^{+0.07}_{-0.04}$ & 0.50$^{+0.04}_{-0.04}$ & 0.99$^{+0.05}_{-0.10}$ & 0.30$^{+0.08}_{-0.08}$ & 1.20$\pm$0.011 & -0.07$\pm$0.011\\
   2434587& 1.80$^{+0.36}_{-0.25}$ & 0.15$^{+0.10}_{-0.12}$ & 1.75$^{+0.40}_{-0.19}$ & 0.30$^{+0.14}_{-0.11}$ & 1.84$^{+0.35}_{-0.25}$ & 0.15$^{+0.07}_{-0.07}$ & 1.78$^{+0.32}_{-0.18}$ & 0.20$^{+0.05}_{-0.05}$ & 1.60$\pm$0.009 & 0.22$\pm$0.001\\
\hline
\end{tabular}

{Ages and metallicities derived from different line-strengths measurements (columns 2-11) for a single SSP  and also inferred from the full spectral fitting technique of {\tt ULySS} (columns 12-13). Errors were estimated with 1000 Monte-Carlo simulations using the errors on the indices and deriving 1$\sigma$ error contours in the age-metallicity space.}               
\end{table*}

\begin{figure}
\centering 
\includegraphics[scale=0.51]{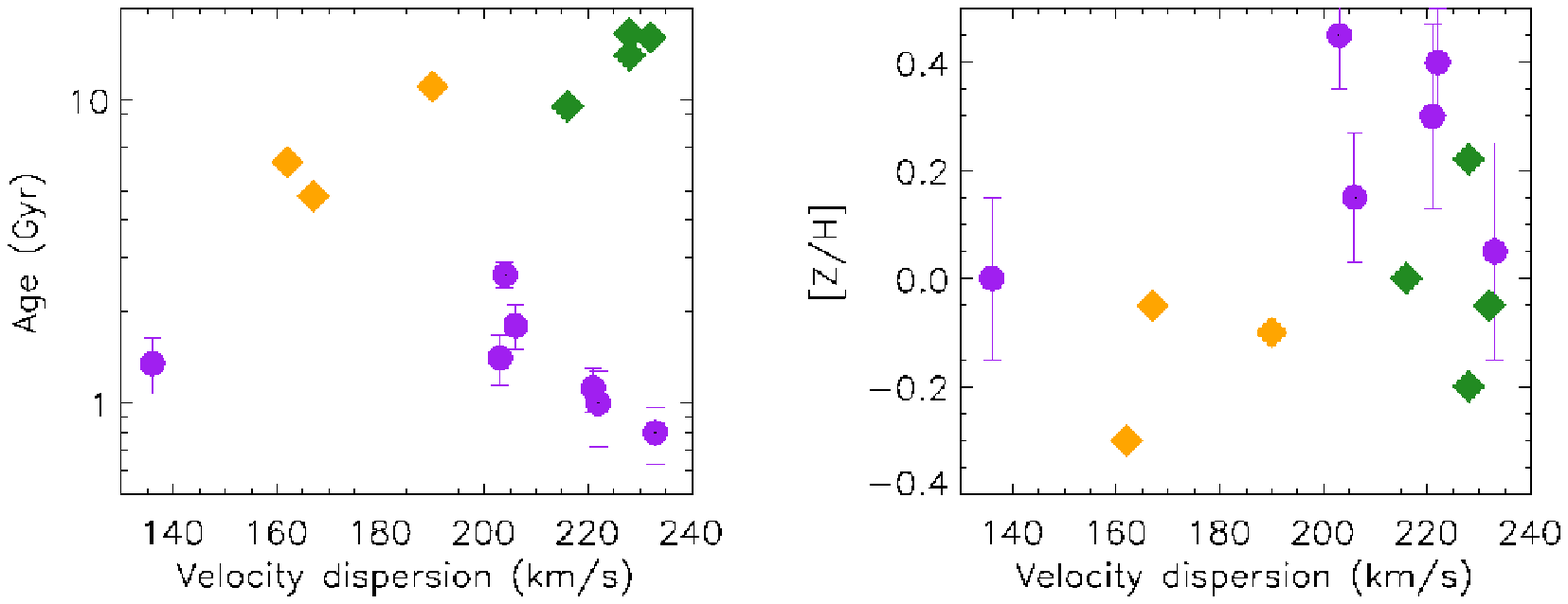} 
\label{Fig.5}
\caption{Derived stellar population parameters plotted \textit{vs} the velocity dispersion for our compact galaxies (purple circles). Control ellipticals (green diamonds) and control spirals (yellow diamonds) are also shown. Note that for a given velocity dispersion, our compact objects do not show similar ages as their local large-sized counterparts, either elliptical or spiral galaxies. Moreover, their total metallicities are systematically richer.}
\end{figure}

\subsection{Gradients}
Taking advantage of the spatial information provided by our high-quality long-slit spectra we also carried out a similar analysis to study possible gradients in the stellar population parameters. The previous analysis was done using a central aperture, but we have extracted two other annular apertures or radii reaching, in a couple of cases, 3\,R$_{e}$ and 4\,R$_{e}$. Table 3 specifies the radial coverage of each aperture, which will be used throughout the paper (r0 for the central aperture, r1 for the first annular aperture, and r2 for the outermost aperture).
Overall, we do not find any strong age or metallicity gradient, as shown in Figure 6. There is a trend suggesting the centers to be slightly younger, but these differences in age are really small ($\leqslant$ 0.5 Gyr). There is also a small trend of the centers to have slightly higher metallicity, but these changes are within the errors. However, gradients in the abundance trends are difficult to interpret because the grids are too degenerated for such low ages, as stated in the previous section.

\begin{table}
\centering
\caption{Annular Apertures coverage}           
\label{table:3}     
\begin{tabular}{c|c c c c }   
\hline\hline      
ID NYU &  r0\,(kpc) & r1\,(kpc) & r2\,(kpc)\\    
\hline  
   54829 &  0.0\,-\,0.9 & 0.9\,-\,1.4 & 1.4\,-\,2.0 \\      
   321479&  0.0\,-\,1.1 & 1.1\,-\,1.9 & 1.9\,-\,2.7 \\     
   685469&  0.0\,-\,1.0 & 1.0\,-\,1.8 & 1.8\,-\,2.5 \\    
   796740&  0.0\,-\,1.5 & 1.5\,-\,2.5 & 2.5\,-\,3.4 \\     
   890167&  0.0\,-\,1.8 & 1.8\,-\,3.0 & 3.0\,-\,4.3 \\      
   896687&  0.0\,-\,1.1 & 1.1\,-\,1.8 & 1.8\,-\,2.5 \\       
   2434587& 0.0\,-\,1.2 & 1.2\,-\,1.6 & 1.6\,-\,2.8 \\    
\hline                                  
\end{tabular}\\
{Radial coverage of each aperture. The notation will we followed through all the paper: r0 for the central aperture, r1 for the first annular radius, and r2 for the second and outermost aperture. }
\end{table}

\begin{figure}
 \hspace{1cm}\includegraphics[scale=0.8]{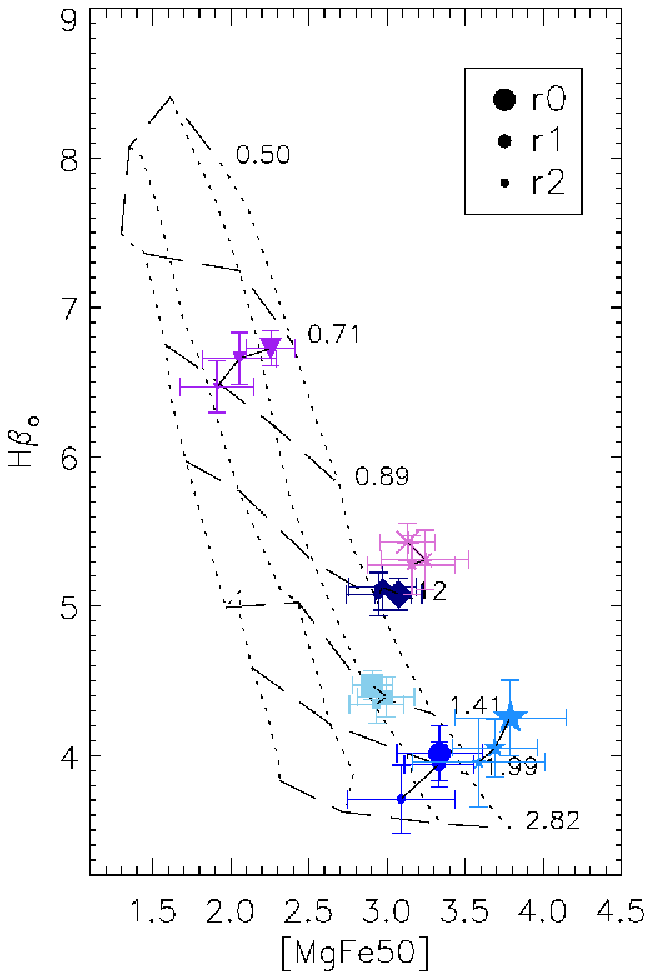}
\label{Fig.6}
\caption{The age-sensitive indicator H$\beta_{o}$ is plotted against the mean metallicity indicator [MgFe50] for our seven compact galaxies. We only show the model grid corresponding to the age range covered by our galaxies. Age (in Gyr) increases from top to bottom as indicated in the labels, and metallicity from left to right ([Z/H]= -0.71, -0.40, 0.00, +0.22). Symbols are like in the previous figures, but the size indicate the different apertures, with decreasing size for increasing aperture. See Table 3 for the corresponding sizes of each aperture on each object.}
\end{figure}

\subsection{Star Formation Histories}
Here we investigate whether the young mean SSP-equivalent ages obtained for these compact galaxies are due to the contribution of recent starbursts that mask an old embeded population, which is dominant in mass, or, alternatively, whether they are genuinely young objects. For this purpose we apply the full spectrum-fitting approach to estimate their Star Formation Histories (SFHs). 
The code used in this section is {\tt STARLIGHT} \citep{CidFernandes2005}, with the V10 SSP SED library. We have defined two bins for the ages:  "young" for lower than the SSP-equivalent age ($\lesssim$\,2\,Gyr) and "old" for $\geq$\,5\,Gyr. We have selected these two bins as the majority of the galaxies do not present populations in between. Table \ref{table:4} (see also Fig.7 and Appendix B) shows the percentages derived from the SFHs. We can separate our objects in two families: (1) those galaxies containing mainly old stellar populations (with a young  contribution lower than 25$\%$ in mass); (2) those with a large fraction of young stars (contributing with more than 25$\%$ in mass). This last result is surprising, as we would not expect these massive galaxies to form nearly half (and even more) of their mass in recent bursts. The uncertainties on the L- and M-weighted "young" parameters were estimated with 2283 pairs of repeated SDSS ETG observations. For all these duplicate spectra, we measured the parameter with {\tt STARLIGHT}, using the same set-up and set of model templates as for the primary spectra. A linear relation is found between the $\Delta$young and the average S/N of each pair. The S/N of the observed spectra are fed into the previous relation to estimate the rms value of the "young" parameters presented on Table 4.\\
The lack of strong gradients derived from the line-strength analysis is confirmed, as fractions remain unchanged through the different apertures. This result also implies that the starburst takes place along the whole structure of the galaxy. \\
Apart from {\tt STARLIGHT}, we repeated the analysis with two other codes, {\tt ULySS} \citep{Koleva2009} and {\tt STECKMAP} (\citealt{Ocvirk2006a}, \citealt{Ocvirk2006b}), all with the V10 SSP SED library. The analysis was repeated by different coauthors independently to double check the possible bias introduced by the selection parameters, the use of different wavelength spectral regions, etc.  We find that the three codes provide very consistent results. In order to trust the recovered SFHs and check the reliability of the results we have also performed a couple of tests, described in Appendix B. The first test constrains the SFH derived with the full-spectrum fittings techniques in terms of the derived fractions. Following the results, we find that all our galaxies show fractions compatibles with their mean SSP-equivalent age (see Fig. B2). The second test checks the robustness of the results, as it compares the mean age and metallicity from the galaxy and the one inferred when measuring the line-strengths of the fitted model (Fig. B3). We are able to recover the same age and metallicity, although the latter with an error of $-$0.10\,dex.

\begin{table}
\caption{Starlight young population fractions from SFHs}           
\label{table:4}     
\begin{tabular}{c|c c c} 
\hline\hline  
ID NYU & Aperture& $\%$young$_{L}$ & $\%$young$_{M}$  \\    
\hline                 
   54829 & r0 & 71$\pm$\,20 & 28$\pm$\,21  \\  
         & r1 & 72$\pm$\,19 & 36$\pm$\,19  \\
         & r2 & 73$\pm$\,19 & 34$\pm$\,19  \\
\hline
   321479& r0  & 70$\pm$\,18 & 17$\pm$\,18  \\ 
         & r1  & 68$\pm$\,17 & 16$\pm$\,17  \\
         & r2  & 68$\pm$\,17 & 16$\pm$\,17  \\
\hline 
   685469& r0  & 82$\pm$\,19 & 65$\pm$\,19  \\
         & r1  & 88$\pm$\,19 & 68$\pm$\,19  \\
         & r2  & 89$\pm$\,19 & 66$\pm$\,19  \\ 
\hline
   796740& r0  & 51$\pm$\,19 & 07$\pm$\,19  \\ 
         & r1  & 52$\pm$\,18 & 07$\pm$\,19  \\
         & r2  & 59$\pm$\,15 & 11$\pm$\,16  \\ 
\hline
   890167& r0  & 99$\pm$\,14 & 97$\pm$\,15  \\   
         & r1  & 98$\pm$\,10 & 88$\pm$\,11  \\
         & r2  & 97$\pm$\,9  & 75$\pm$\,10  \\
\hline 
   896687& r0  & 60$\pm$\,18 & 12$\pm$\,18  \\  
         & r1  & 55$\pm$\,15 & 10$\pm$\,16  \\
         & r2  & 55$\pm$\,14 & 10$\pm$\,15  \\ 
\hline
   2434587& r0 & 60$\pm$\,20 & 22$\pm$\,20  \\   
          & r1 & 62$\pm$\,19 & 23$\pm$\,19  \\
          & r2 & 68$\pm$\,18 & 35$\pm$\,19  \\ 
\hline                                  
\end{tabular}\\
{Fractions of young ($\lesssim$2\,Gyr) stellar populations retrieved from the SFHs of the galaxies using the {\tt STARLIGHT} full-spectral-fitting code. First, the luminosity-weighted fractions, second the mass-weighted ones. For each galaxy, the first line corresponds to the central aperture, the second to the second aperture and the third to the outermost radius considered for this study. }
\end{table}

\begin{figure*}
  \begin{center}
    \begin{tabular}{cc}
        \includegraphics[scale=0.45]{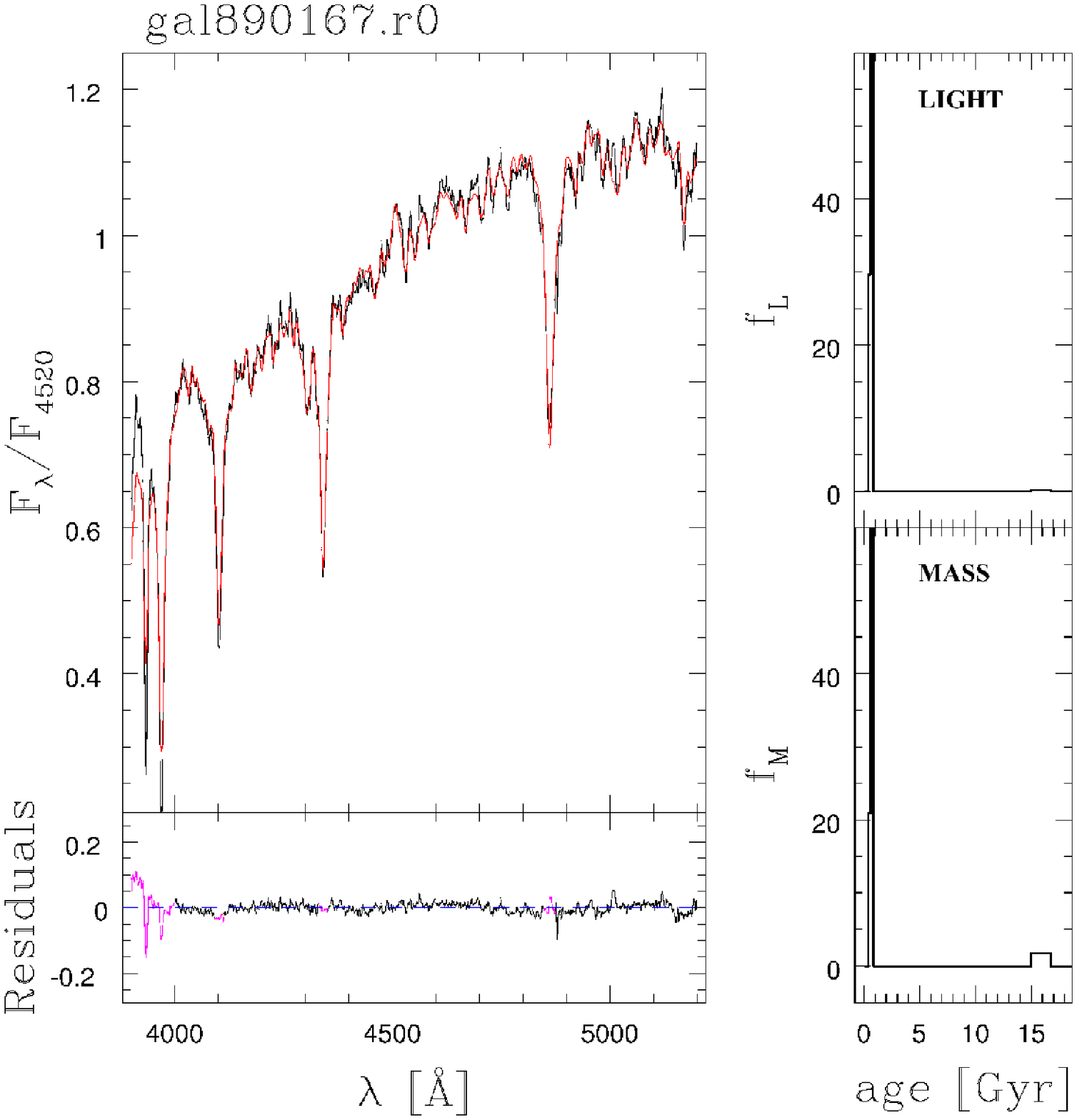} &
        \includegraphics[scale=0.45]{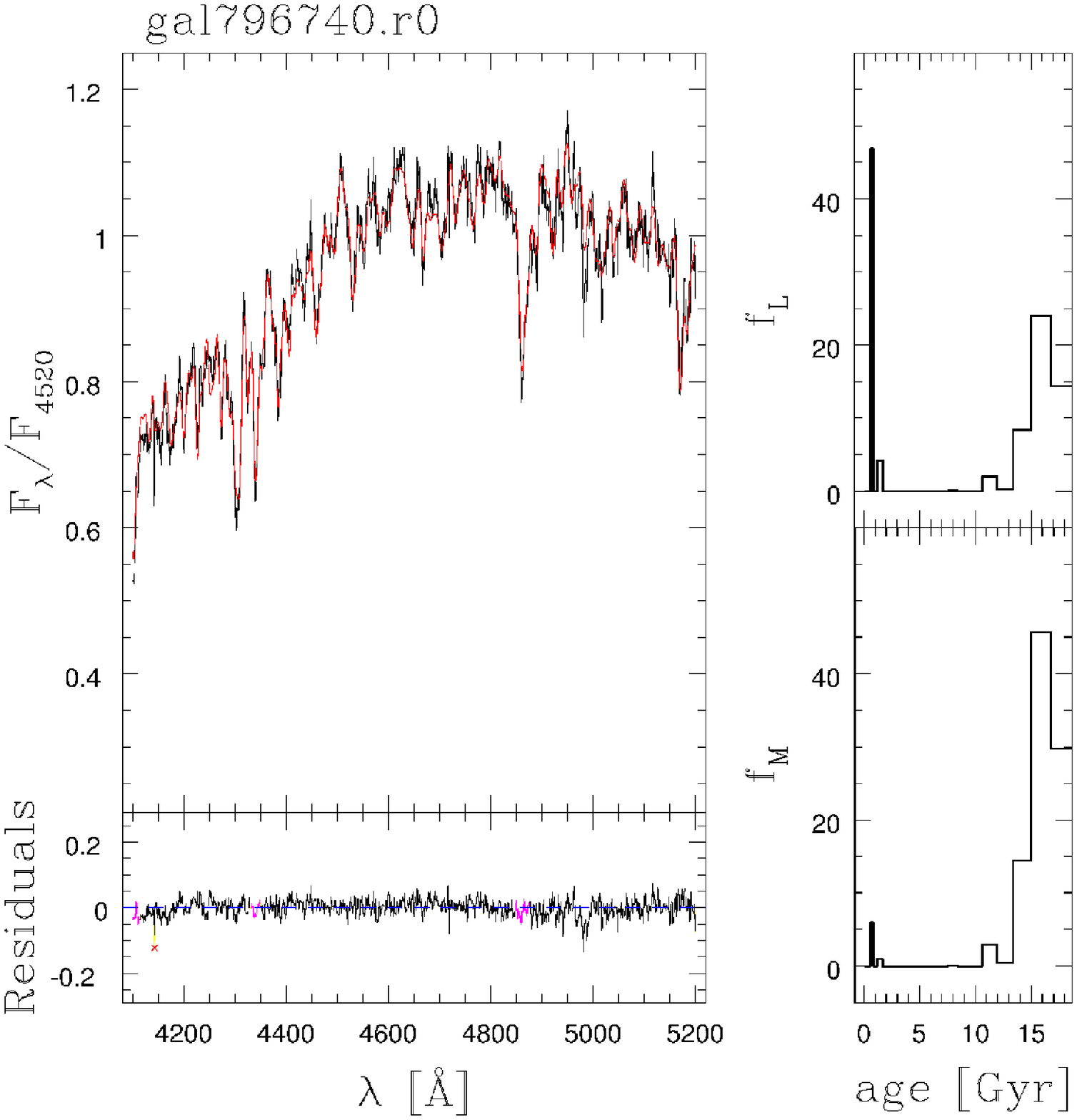}
      \end{tabular}
\label{Fig. 7}
\caption{Results from {\tt STARLIGHT} for the central apertures of galaxies 2434587 and 54829. On the upper-left hand, the galaxy spectra (black) is fitted with the models from V10 (red), bottom-left panel shows the fitting residuals. The recovered Star Formation History are also shown, in light (upper-right panel) and in mass (down-right panel) for each galaxy. All our objects show recent bursts, some of them leading to a big contribution from the young stellar populations. The rest of galaxies SFHs can be found in Appendix B.}
  \end{center}
 \end{figure*} 

\section{Mass Estimates}
Finally, we have revised the mass estimates for these objects, as the mass was one of the selection criteria of the original sample. We have estimated the stellar and dynamical masses following different approaches, as described below. The results are summarized in Table \ref{table:5}.\\
As mentioned before, our galaxies were selected from the NYU VAGC (B07) with a mean stellar mass for these seven galaxies of M*$_{B07}\sim$\,9.2\,$\times$\,10$^{10}\,M_{\sun}$. However, the catalog from the SDSS/DR7/MPA (http://www.mpa-garching.mpg.de/SDSS/DR7/; hereafter MPA), gives systematically lower masses, with a mean stellar mass of M*$_{MPA}\sim$\,8.3\,$\times$\,10$^{10}\,M_{\sun}$. The mass estimates of these two analysis were obtained by fitting the broad-band SDSS (\textit{u,g,r,i,z}) photometry, with the models of \citet{Bruzual2003} (BC03, hereafter) and a Chabrier IMF (Initial Mass Function). However, in B07 they added the near-IR photometric bands of 2MASS (\textit{J, H, K$_{s}$}). Following the approach of \citet{Kauffmann2003a} based on the spectral fitting, instead of the photometry, we used the SSP model SEDs of V10 with a Kroupa Universal IMF. For this purpose we use the M/L corresponding to the SED model that best fits the galaxy spectra, which provides a mean luminosity-weighted age and metallicity. The stellar mass was inferred by calculating the M/L in the SDSS r-band, adopting the {\tt GALFIT} magnitude as luminosity estimate, after correcting for galactic extinction. The stellar masses computed in this way take into account the mass of the luminous stars and the remnants which, for the ages we are considering ($\sim$ 2\,Gyr), is of the order of 0.6 times the initial mass of gas converted into stars. The mean stellar mass calculated this way is M*$_{SSP}\sim$\,4.3\,$\times$\,10$^{10}\,M_{\sun}$.
In a different approach, we consider instead of an equivalent-SSP stellar population, the spectra resulting from integrating the contribution of different SSPs, with different M/L, according to the SFH derived with {\tt STARLIGHT} (see Section 4.3), leading to a mean M*$_{SFH}\,\sim$\,1.5\,$\times$\,10$^{11}\,M_{\sun}$. For comparison, we carried a similar approach with the SDSS fiber spectra, but with a longer wavelength-range, using the mass-weighted SFHs solution from {\tt STARLIGHT}, obtaining a mean stellar mass of M*$_{SDSS}\,\sim$\,1.3\,$\times$\,10$^{11}\,M_{\sun}$.  \\
From the above estimates it is clear that the stellar mass determination is not straightforward. It has long been reported that the methodology, the models, the selection of the IMF and other parameters, strongly affect these determinations (see e.g.: \citealt{Bell2001}, \citealt{Kauffmann2003a}, \citealt{Swindle2011} and \citealt{Cappellari2012}). For example, B07 gives a mean mass of  M*$_{B07}\sim$\,9.21\,$\times$\,10$^{10}\,M_{\sun}$ for a Chabrier IMF that translates to M*$_{B07}\sim$\,1.0\,$\times$\,10$^{11}\,M_{\sun}$ when employing a Kroupa, more similar to our M*$_{SFH}$. 
The masses derived with the SFHs are systematically larger than the equivalent-SSP measurements. Since in the SSP-method, the ages are luminosity weighted, the final values are strongly biased to the youngest stellar component present in the galaxy and, therefore, the M/L ratios are also biased towards lower values, as their stellar masses. The SFH derived from full-spectrum fitting instead, reveals an important presence in terms of mass fractions of old stellar populations, whose M/L is considerably higher than those of the SSP corresponding to the luminosity-weighted age. For this reason, the M/L based on SFH tend to be larger than the ones derived with a single stellar population. Indeed, the objects with the largest contribution from old stellar populations (321479, 796740 and 896687 and 2434587) are the ones showing the largest differences with respect to the masses derived by means of the SSP-equivalent parameters (by a factor 4). Instead, the three galaxies with the highest fraction of young stellar population show a more modest difference between the two estimates (by a factor of 2). We have tested the M/L obtained from the SFHs derived with two full-spectral-fitting codes, {\tt STARLIGTH} and {\tt STECKMAP}, finding consistent results. As most of the results presented here come from  {\tt STARLIGTH}, we use the masses computed from their SFHs. Moreover, we have calculated the M/L inferred by a single SSP and by the SFH on our control elliptical galaxies. As their luminosity and mass-weighted ages are similarly old, we should not find a relevant difference in the M/L, as we do.\\
The dynamical mass is considered to be a good proxy for the mass of a galaxy. The estimate comes from the Scalar Virial Theorem (\citealt{Binney2008}, and references therein). As our objects show both velocity dispersion and rotation motions (the latter to a lesser extent), we may account for both contributions for the dynamical mass \citep{Epinat2009}: M$_{dyn}$=\,M$_{\theta}$+\,M$_{vir}$. These contributions are shown in Table 5. The rotationally supported term is the one contributing less for our galaxies:
\begin{equation}
\centering
 M_{\theta} = [V^{2} R]/G,
\end{equation}
where V is the velocity at radius R and \textit{G} the universal gravitational constant. We choose V=\,V$_{max}$ for those galaxies for which we reached the maximal velocity rotation, while we take V=\,V$_{high}$ for the rest, considering the highest radial velocity obtained. The three galaxies showing the highest radial velocities (from Sect. 3.2) are the ones showing a larger contribution from this term. 
Following the approach from \citet{Bertin2002}, the virial mass is:
\begin{equation}
\centering
 M_{vir,n} = [K(n)\sigma^{2} R_{e}]/G,
\end{equation}
where
\begin{equation}
\centering
 K(n) \simeq \frac{73.32}{10.465+(n-0.95)^{2}}+0.954,
\end{equation}
where \textit{n} is the S\'ersic index, \textit{R$_{e}$} the circularized effective radii in kpc and $\sigma$ the velocity dispersion in $\rm{km\,s^{-1}}$. This \textit{K(n)} factor takes into account the geometry of the mass distribution and velocity projection along the line of sight. Under the assumption of homology, Cappellari et al. (2006) determined that for a de Vaucouleurs r$_{1/4}$ profile, this factor is equal to 5. Table \ref{table:5} shows the dynamical masses under the assumptions of either homology and non-homology. We obtain a mean M$_{dyn(nohom)}$\,=\,6.89\,$\times$\,10$^{10}\,M_{\sun}$ and M$_{dyn(hom)}$\,=\,7.65\,$\times$\,10$^{10}\,M_{\sun}$, both lower than the stellar masses. The mass estimates for the galaxy 54829 are not taken into account to calculate mean masses, as it is the only one having a much lower $\sigma$ (that contributes quadratically on the equation), which would strongly bias the mean values to lower masses. For consistency, we neither use this galaxy for the stellar mean masses.

Figure 8 shows different stellar masses \textit{versus} the dynamical mass. Filled symbols refer to the homology virial mass, open symbols to the non-homology one. The dashed line corresponds to the 1:1 relation. The first noticeable result is the difference in the stellar mass depending on the methodology adopted, as stated above. The lowest stellar masses are obtained with the SSP approach (M*$_{SSP}$, yellow circles), and they increase when using the SFHs derived for the galaxy (M*$_{SDSS}$, green stars). In between, we have the stellar masses from the photometric fitting in B07 (violet diamonds). However, the most intriguing point is that the derived stellar masses are larger, on average, than the dynamical masses (except for the M*$_{SSP}$). Previous studies have shown that dynamical masses are on average equal or larger than stellar masses (\citealt{Cappellari2006}, \citealt{vanderWel2006}, \citealt{vandeSande2011}). However, the study of four compact and massive early-type galaxies at z$\sim$\,1 from \citet{Martinez-Manso2011} showed that the dynamical masses where smaller, by an average factor of $\sim$\,6, although they could not find a plausible explanation. Where does the problem come from? Are the stellar masses wrong or are the dynamical ones? Or both? \\ 
On one hand, regarding to the stellar mass, the discrepancies may be related to the large contribution of young stellar populations, as discussed above. The SSP-equivalent stellar masses are in much better agreement with the dynamical masses, as the two are mainly dominated by the young stellar components and the comparison is consistent (dynamical masses are mainly determined by the velocity dispersion of the galaxy, which might be biased by the young populations that dominate the light). Instead, the SFH-based stellar masses are in worse agreement and larger than the dynamical masses, as the old stellar components present in the derived SFH lead to an inconsistent comparison. Moreover, the recent paper of \citet{Cappellari2012} has shown that variations of the IMF with the mass of the galaxy should be taken into account, being then another source in the discrepancy. On the other hand, if we consider that the problem is on the dynamical mass, we have to draw special attention to the sizes of our objects. As stated in \citet{Stockton2010}, it could be that we are using a definition for the dynamical mass that was initially calibrated for normally-sized objects, while our galaxies are extremely compact. Should we double our galaxy sizes, we would obtain higher dynamical masses, then we would be closer to the 1:1 relation. In a forthcoming paper of Trevisan et al. (in prep.), the authors present the M$_{dyn}$-M* relation for 40000 ETGs from the SDSS \citep{LaBarbera2010}, showing a large fraction of galaxies with stellar masses greater than their dynamical masses. Moreover, they show that those galaxies, always present small sizes, while they show a whole range in ages. These results point out to the dynamical mass to most likely be the wrong one.\\

\begin{table*}
\caption{Galaxy Mass Estimates}           
\label{table:5}     
\centering                      
\begin{tabular}{c|c c c c c c c c c c}   
\hline\hline      
ID NYU & M*$_{SSP}$ & M*$_{SFH}$ &  M*$_{SDSS}$ & M*$_{B07}$ & M*$_{MPA}$ & M$_{v}$& M$_{v,n}$ & M$_{\theta}$& M$_{dyn(h)}$& M$_{dyn(noh)}$  \\
& & & & &[$\times$\,(10$^{10}M_{\sun}$)]& & & & & \\    
\hline  
   Model& V10 & V10 & V10 & BC03 & BC03 & V10 & V10 & V10 & V10 & V10 \\ 
   IMF &  Kroupa & Kroupa & Kroupa & Chabrier & Chabrier& Kroupa & Kroupa & Kroupa & Kroupa & Kroupa \\ 
   type & spectra & spectra & spectra & phot & phot & spectra & spectra & spectra & spectra & spectra \\ 
   range(nm) & [380-530] & [380-530] & [360-735] & SDSS+2MASS & SDSS & [380-530] & [380-530]& [380-530]&[380-530] & [380-530] \\ 
\hline
   54829  & 4.06 & 9.61 & 8.22  & 8.39 & 7.09 & 2.56 & 1.91 & 0.04 & 2.60 & 1.95 \\      
   321479 & 5.01 & 20.30& 16.16 & 10.10& 8.19 & 6.81 & 3.52 & 1.78 & 8.59 & 5.30 \\
   685469 & 4.67 & 5.45 & 11.10 & 9.03 & 7.29 & 7.23 & 8.54 & 0.38 & 7.61 & 8.92 \\
   796740 & 4.45 & 23.05 & 13.72 & 9.76 & 7.65 & 6.05 & 8.06 & 2.32 & 8.37 & 10.38\\
   890167 & 2.05 & 3.21 & 13.92 & 8.79 & 12.49& 5.11 & 5.11 & 2.10 & 7.21 & 7.21 \\
   896687 & 5.11 & 23.12 & 11.18 & 9.23 & 7.39 & 8.13 & 5.44 & 0.02 & 8.15 & 5.46 \\
   2434587& 4.31 & 14.42 & 10.63 & 8.37 & 6.77 & 5.52 & 3.69 & 0.43 & 5.95 & 4.12 \\
\hline   
   MEAN  & 4.27 & 14.89 & 12.78 & 9.21 & 8.29 & 6.47 & 5.72 & 1.17 & 7.65 & 6.89 \\
\hline                                  
\end{tabular}\\
{Stellar and dynamical mass estimates. M*$_{SSP}$ represents the SSP-equivalent mass estimate, while M*$_{SFH}$ refers to the one computed with the mix of SSPs from the derived SFHs of each galaxy (see sect. 4.3), for our spectra (M*$_{SFH}$) and for the SDSS spectra (M*$_{SDSS}$). M*$_{B07}$ and M*$_{MPA}$ are the values from B07 and MPA catalogs. For the dynamical masses, the virial masses are M$_{v,n}$ and M$_{v}$, the first accounting for the non-homology. The rotation contribution is stated as M$_{\theta}$ and M$_{dyn(h)}$ and M$_{dyn(noh)}$ are the total dynamical masses again accounting for the homology/non-homology (\citealt{Cappellari2006}). All  mean mass values were computed without the galaxy 54829, due to its low velocity dispersion in comparison to the other galaxies.}
\end{table*}

\begin{figure}
  \includegraphics[scale=0.8]{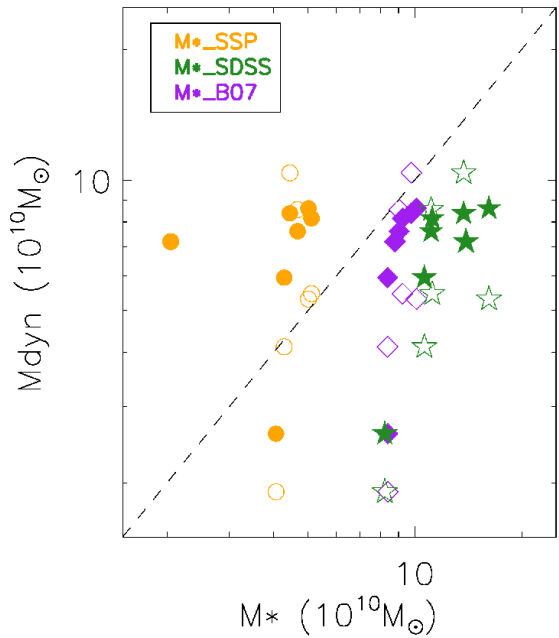} 
\label{Fig.8}
\caption{Dynamical mass \textit{vs.} stellar mass computed with different approaches. Filled symbols state for the homology (\textit{K(n)=5}), while open symbols state for the non-homology. Yellow circles are the M*$_{SSP}$, violet diamonds are M*$_{B07}$ and green stars are M*$_{SDSS}$. The dashed line is the 1:1 relation. Mainly all the stellar masses, except for the M*$_{SSP}$, are similar or bigger than the dynamical mass.}
 \end{figure}

\section{Discussion and Conclusions}
In this work we present the main properties of 7 of the 29 bonafide compact massive galaxies identified by T09 in the nearby Universe (z\,$\leq$\,0.2). We find that these galaxies represent a peculiar type of objects, not only from a detailed analysis of their morphologies and stellar kinematics, but also according to their stellar populations. The main aim of this work is to characterize these objects in order to understand better their formation and evolution. We measure large radial velocities (up to 200\,$\rm{km\,s^{-1}}$) and high velocity dispersions ($\sim$\,200 $\rm{km\,s^{-1}}$) for these objects. Morphologically, they are very compact ($Re\,\sim$\,1.3\,kpc) and with S\'ersic indices typically higher than 3, indicative of spheroids. However, according to their axial ratio, they look elongated on the sky (\textit{b/a}$\la$\,0.6). From an analysis of their stellar populations, we derive young mean SSP-equivalent ages ($\leqslant$\,2\,Gyr) and metallicities solar or higher ([Z/H]\,$\geq$\,0). The full-spectrum-fitting derived SFHs reveal that these objects contain an old embeded population but that they all experienced recent bursts of star formation during the last 2\,Gyr. The latter contribute strongly to the total light ($\ga$\,50$\%$) and, in some galaxies, even to the total mass ($\ga$\,30$\%$). We divide our sample in two classes depending on the strength of the recent bursts: (1) Four galaxies show mainly old stellar populations ($\geq$\,75$\%$ of the total stellar mass); (2) and three objects contain a very important contribution of young stellar populations ($\ga$\,30$\%$).
 
Although local compact galaxies show velocity dispersions and S\'ersic indices similar to those in local normally-sized massive ellipticals, they show important rotation curves and seem to be rotationally supported. Moreover, unlike ellipticals of similar mass, our compact galaxy sample shows young SSP-equivalent ages ($\leq\,$2\,Gyr) and higher metallicities.\\
Despite the fact that our galaxies display disky morphologies and strong rotation curves, they differ from spiral galaxies in their S\'ersic indices and their velocity dispersions (the latter showing lower values for both; \textit{n}\,$\leqslant$\,2.5 and $\sigma\,\leqslant$\,150\,$\rm{km\,s^{-1}}$). Moreover, our objects are younger than spirals of similar mass.\\
For completeness, we compare our objects with dwarf ellipticals, although their masses are much lower than the ones measured in our sample. Many dwarf galaxies show young mean luminosity-weighted ages, which can be similar to those inferred for our galaxy sample, but in general they have lower metallicities. Kinematically, our galaxies are rotating as dwarf ellipticals, but the latter have lower S\'ersic indices and lower velocity dispersions. \\
Our objects present optical spectra resembling 'E+A' galaxies (\citealt{Dressler1983},
\citealt{Couch1987}). E+A objects are considered post-starburst galaxies due to the presence of strong Balmer absorption lines, indicative of a young population ($<$1\,Gyr) and an absence of major emission lines ([OII] or H$\alpha$) indicating the cessation of star formation. It is commonly accepted that E+A represent the transition phase from a disky, rotationally supported system to a spheroid, pressure supported one (\citealt{Yang2004}, \citealt{Zabludoff1996}, \citealt{Pracy2009}), showing a widespread of properties (ages, metallicities, velocity dispersions, sizes).
Because of their optical spectra and their SFHs, our objects are post-startburst galaxies. However, it is worth noticing that none of our galaxies, which we selected from the SDSS database, was previously identified as E+A (e.g., see the catalogs of \citealt{Blake2004}, \citealt{Goto2005}, \citealt{Goto2007a}). The different catalogs take different limiting values on the equivalent width of the lines that determine the family, reason why our objects may not be found in them. \\

The goal of this work was to characterize a sample of local compact massive galaxies, which were initially thought to be the relics of massive compact objects at high-z. Instead, we have shown here that they seem to be exact copies of those high-z galaxies, as their inferred properties (masses, sizes, luminosity-weighted ages) are nearly the same. We must note, however, that their SFHs do not agree with the massive compact galaxies at high-z, which are genuinely young at those redshifts. Although our objects have experienced important recent starbursts, some of them are compatible with having very old stellar populations. We compared the properties of our sample with those of ellipticals and spirals in the local Universe. While they share a number of similarities, the compact galaxies also present many important differences (e.g., their young ages) and they deviate from various relations expected in the local Universe. These differences do not allow us to classify them neither as ellipticals nor spirals. Instead, their optical spectra resembles those of the E+A galaxies. May it be, that we are in fact seeing an intermediate stage on the thought transformation from spirals to ellipticals via E+As? We have found a new class of objects, unique and rare in the nearby Universe.\\
Based on the derived SFHs, we can divide our objects in two groups. The first one is composed by those compact massive galaxies with large fractions of old stellar populations. These objects could be considered the relics of the high-z universe, although their SSP-ages are still young, due to the non-negligible fraction of young components. In the second group, we include those galaxies with a huge fraction in mass of young stellar populations, analogous to the high-z massive galaxies. This group is the most intriguing. How is it possible to form half of the mass of a massive galaxy so recently ($\le$\,2\,Gyr)? Whatever mechanism responsible for this, demands huge quantities of gas to create this amount of stars in such a short time. Is this mechanism analogous to that forming the massive galaxies at high-z (cooling flows or the merger of two ``gas rich`` spirals, e.g. \citealt{Ricciardelli2010})? The absence of clear stellar gradients along the nearby compact galaxies suggests that the mechanism triggering the starburst in these objects is likely a global one. This could indicate that the structure of the nearby compact massive galaxies has been assembled during this major event, being the old stellar populations the stars associated to the merging units and the young component a product of the gas collapse into stars. \\
Although we have characterized their stellar populations and kinematics, we still need to understand their formation and evolution and assess the issue of why these local massive galaxies were born with such a compact structure. A first approach to solve these open questions would be to expand our wavelength range as, for the moment, only optical data is available. Moreover, we will have to wait until high-quality spectra for the high-z analogs are obtained, to compare their relevant stellar properties (e.g., the abundance ratios and total metallicity). We also require high quality imaging to properly look to their morphologies and trace their merging activity (see e.g. \citealt{Trujillo2012}).\\
The local compact galaxies analyzed in this paper seem to be unique and scarce in the nearby Universe and, because of their relative vicinity, they are a perfect laboratory to explore with unprecedented detail the formation mechanisms of massive galaxies at high-z, particularly those involving huge amount of gas leading to the enormous formation of stars we have measured.\\

\section*{Acknowledgments}
We thank the anonymous referee for a careful reading of the paper and for making detailed
suggestions that improved the understanding of this work. We thank J. Falc\'on-Barroso for his help on using the new MILES webpage and for constructing the spiral control sample, and V. Quilis for useful discussion. Based on observations made with: WHT/ING Telescope in El Roque the los Muchachos Observatory of the Instituto de Astrof\'isica de Canarias, La Palma. This work has been supported by the Programa Nacional de Astronom\'ia y Astrof\'isica of the Spanish Ministry of Science and Innovation under grant AYA2010-21322-C03-02. PSB is supported by the Ministerio de Ciencia e Innovaci\'on (MICINN) of Spain through the Ram\'on y Cajal program. PSB also acknowledges a Marie Curie Intra-European Reintegration grant within the 6th European framework program and financial support from the Spanish Plan Nacional del Espacio del Ministerio de Educaci\'on y Ciencia (AYA2007-67752-C03-01). ER is supported by the Programa Nacional de Astronom\'ia y Astrof\'isica of the Spanish Ministry of Science and Innovation under grant AYA2010-21322-C03-01.

\bibliography{compactas2}
\bibliographystyle{mn2e}

\appendix
\section{Galfit}
In this section we show the rest of our galaxies not shown in Section 3. First panel corresponds to the original image from the SDSS database, the second is the model created with the PSF of a suitable star close to our object, and the third panel shows the residuals from the substraction. We chose these as the best results by their reduced $\widetilde{\chi^{2}}$ and by visual inspection. It is clear that for some galaxies there is a remaining residual in the center, which could not be removed with a simple S\'ersic profile. Bulge-disk decompositions could be made, but SDSS quality is not optimal, high-quality images are needed to properly study the morphologies of these objects.

\begin{figure*}
 \includegraphics[scale=1.0]{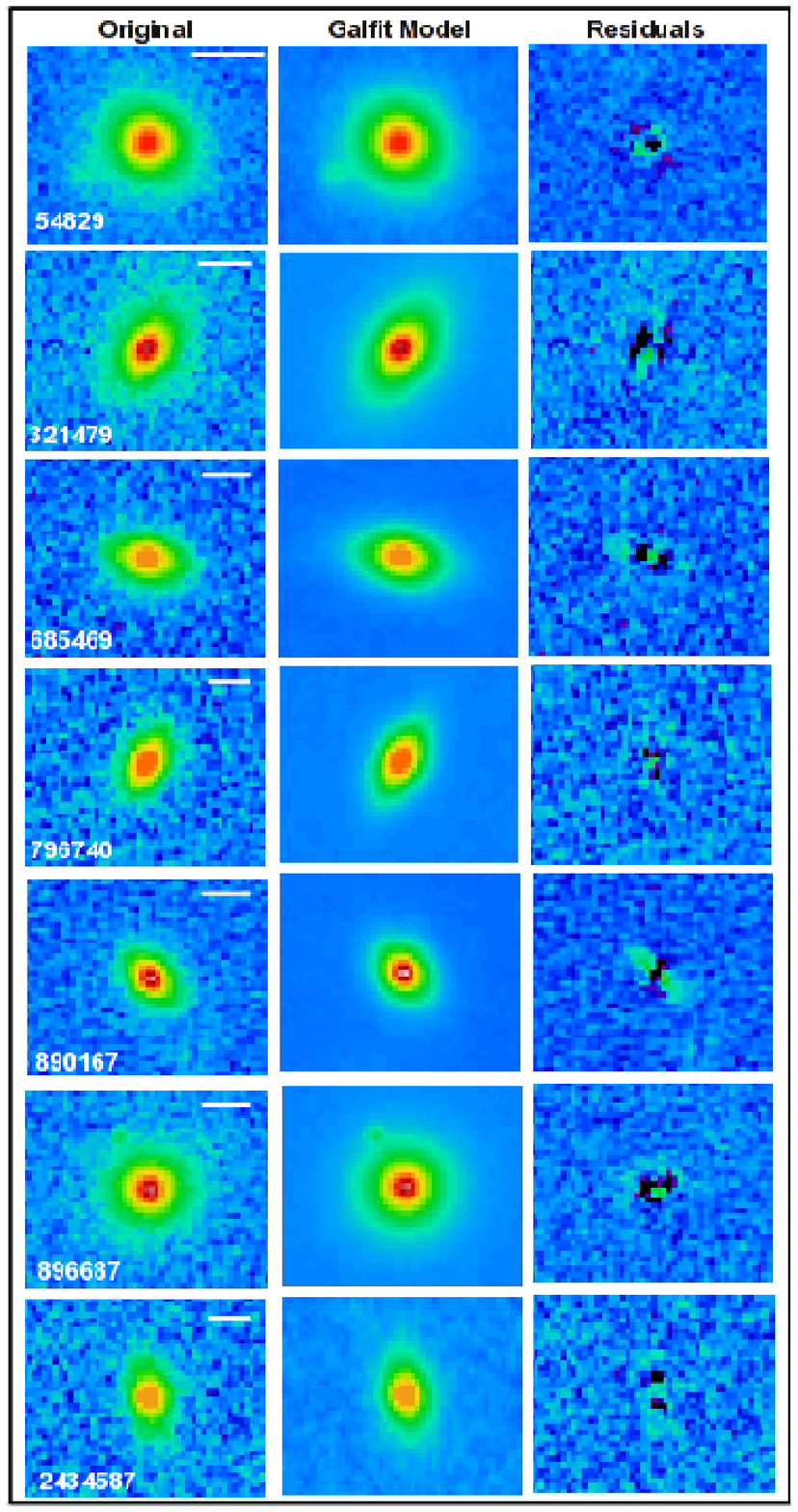}
\caption{Thumbnails for each compact massive galaxy showing for each object, the galaxy image from SDSS (first panel), the fitted model with {\tt GALFIT} (second panel) and the residuals from the fit (third panel). The white line on the upper right corner of all the first panels is the equivalent size of 10\,kpc at the redshift of the object.}
 \end{figure*} 

\section{SFHs}
In this Appendix we show the SFHs recovered with the code {\tt STARLIGHT}. Each full-spectrum-fitting code gives us different information, so we choose the most suitable in each case. For example, {\tt ULySS} is very efficient giving 1-SSP fits, as shown in Table 2. However, we find more useful the output from {\tt STARLIGHT} when studying the contribution of the main stellar populations. Finally, {\tt STECKMAP} was used as double-check for our results.  All three codes were fed with the V10 models.
For each galaxy, we have analyzed the SFH for the different apertures, to study possible gradients. In Fig. B1, each row on the panel corresponds to a single galaxy, with the first column being the center (r0), the middle one the first aperture (r1), and the third one the outermost aperture (r2). Each panel contains the galaxy spectra (in black), the fitted spectra from the model templates (red) and the residuals, showing, if any, the rejected pixels and masked regions (green). The SFHs are shown on the right of each spectra, both in luminosity (up) and mass (down).

\begin{figure*}
        \includegraphics[scale=0.28]{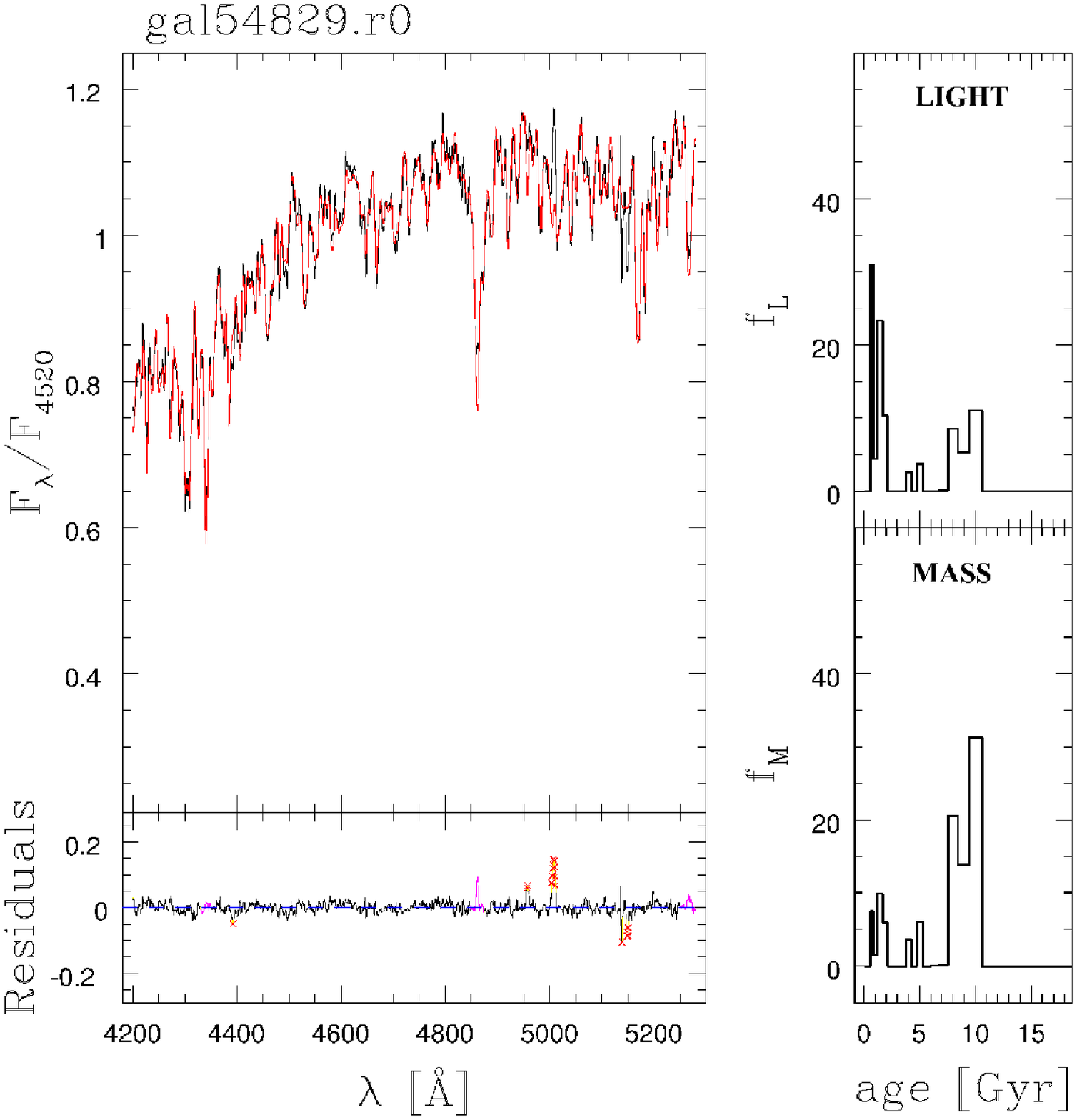}  
        \includegraphics[scale=0.28]{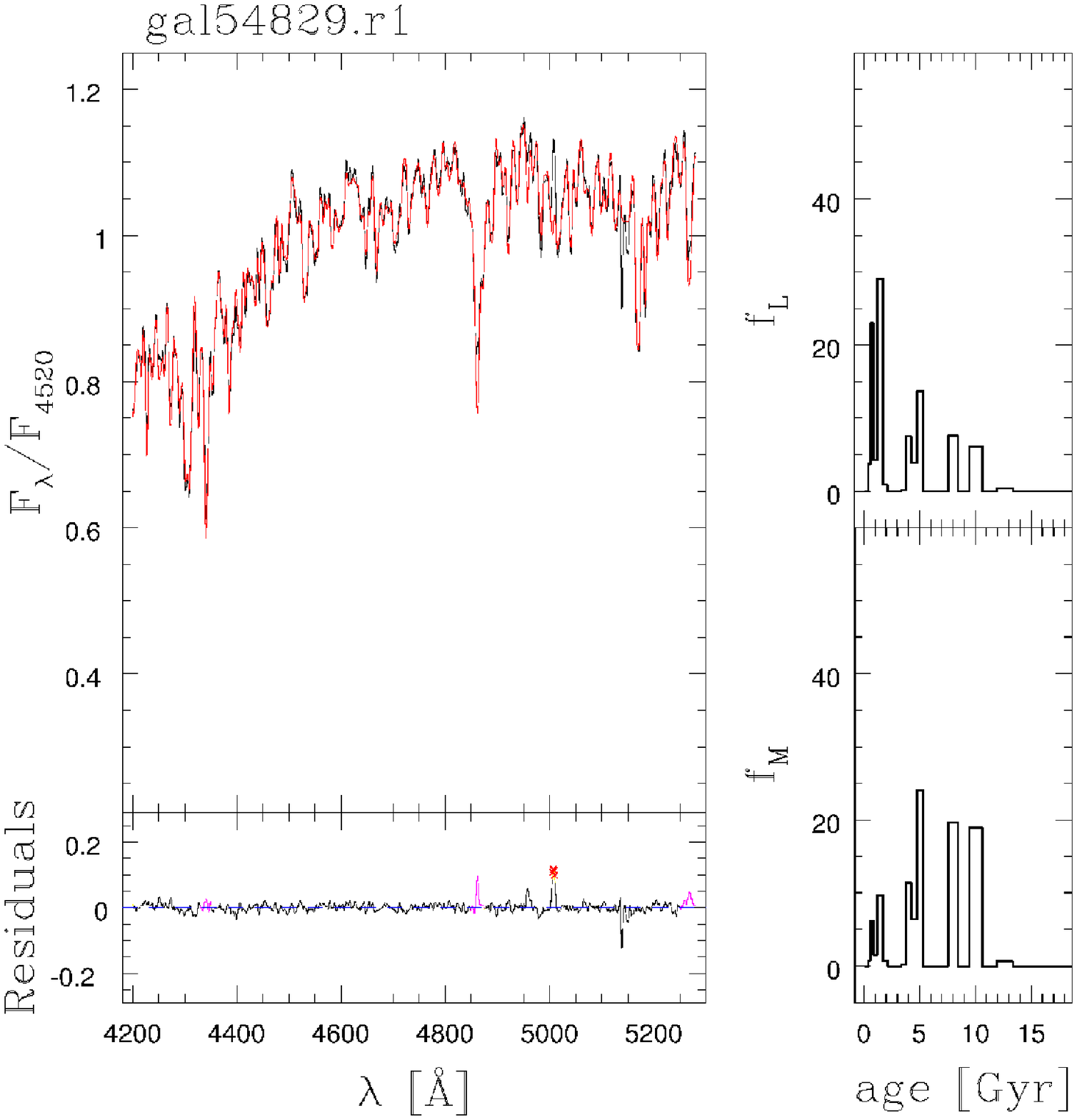} 
        \includegraphics[scale=0.28]{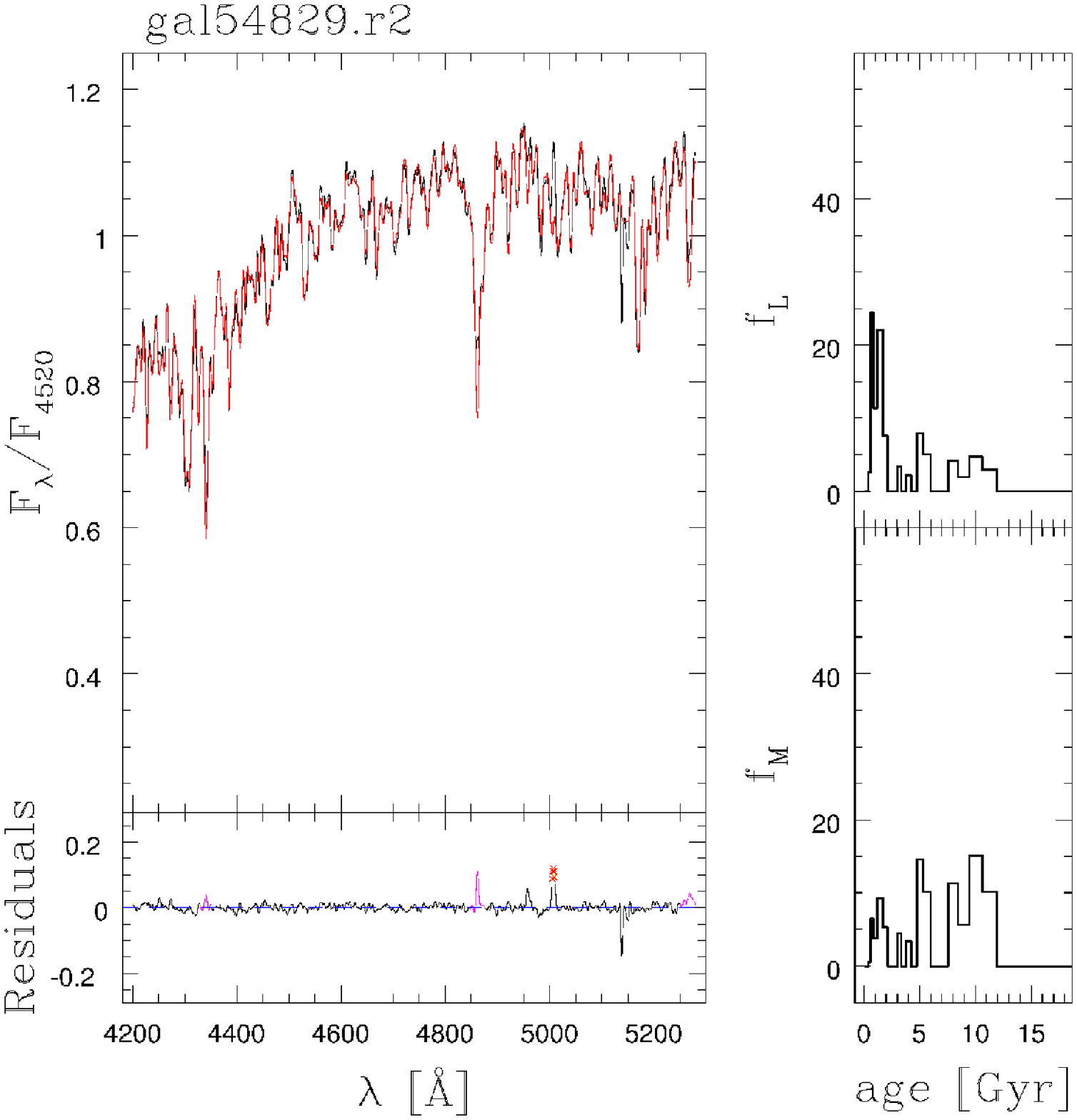} \\
        \includegraphics[scale=0.28]{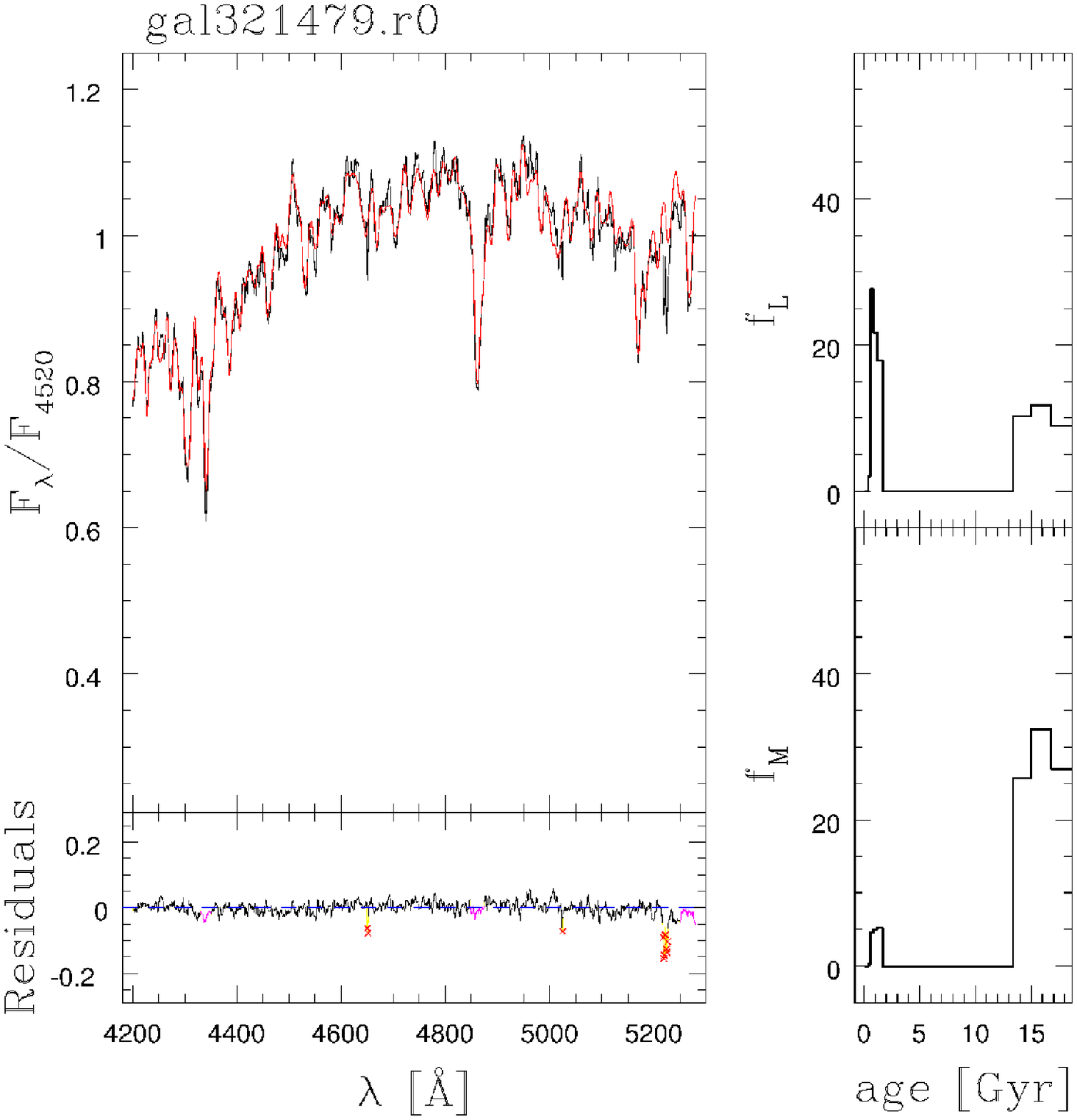} 
        \includegraphics[scale=0.28]{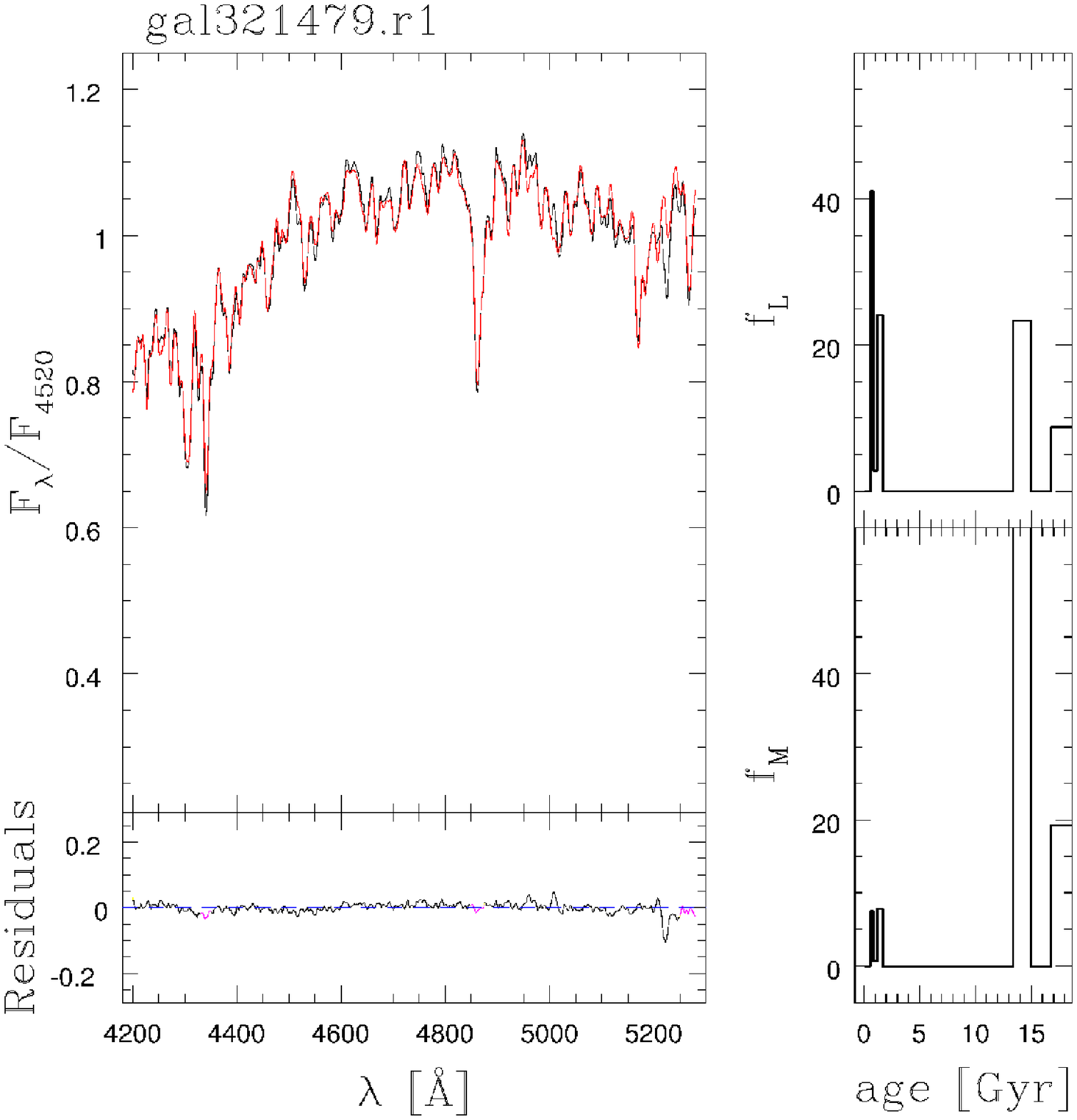} 
        \includegraphics[scale=0.28]{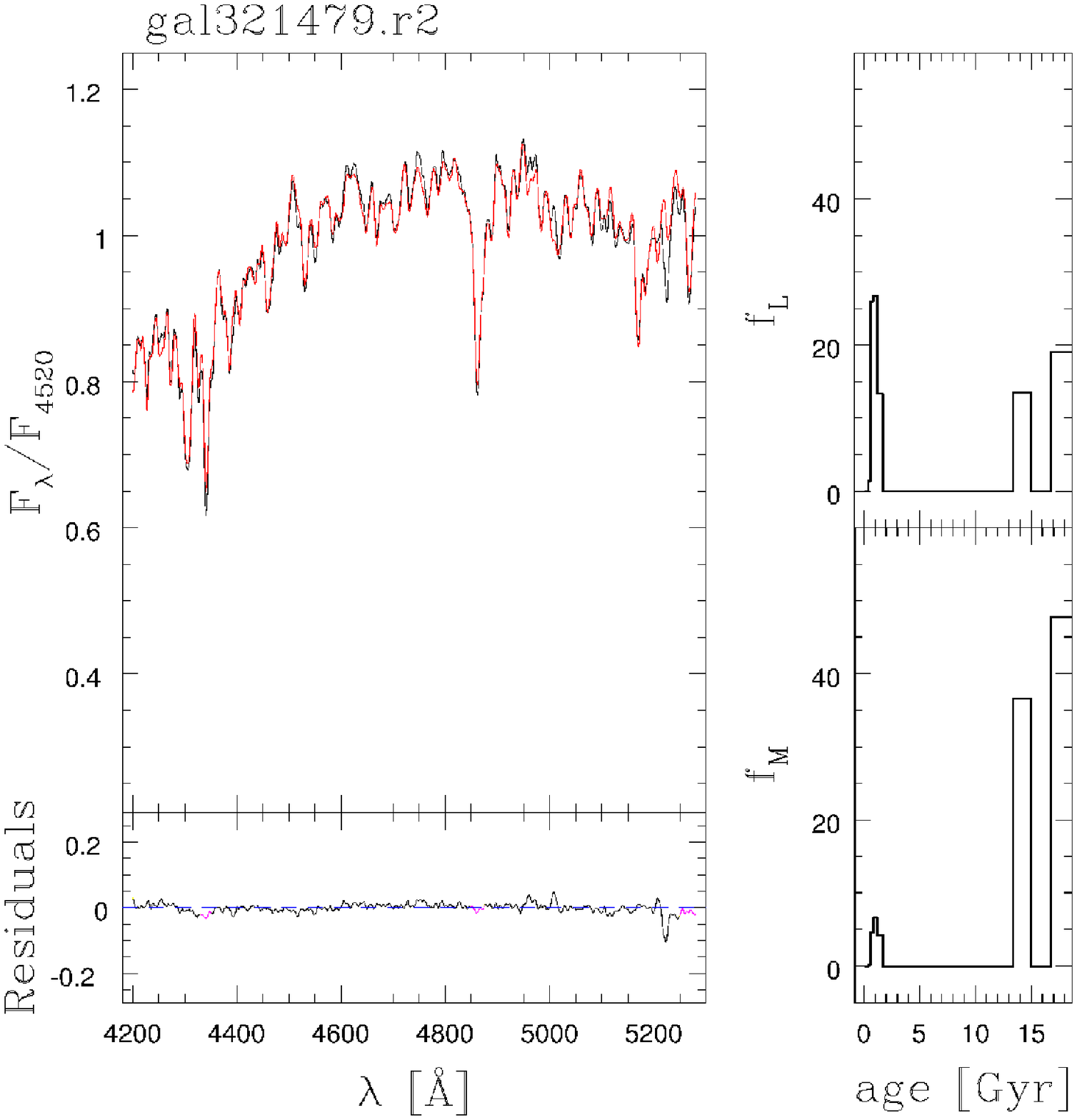} \\
        \includegraphics[scale=0.28]{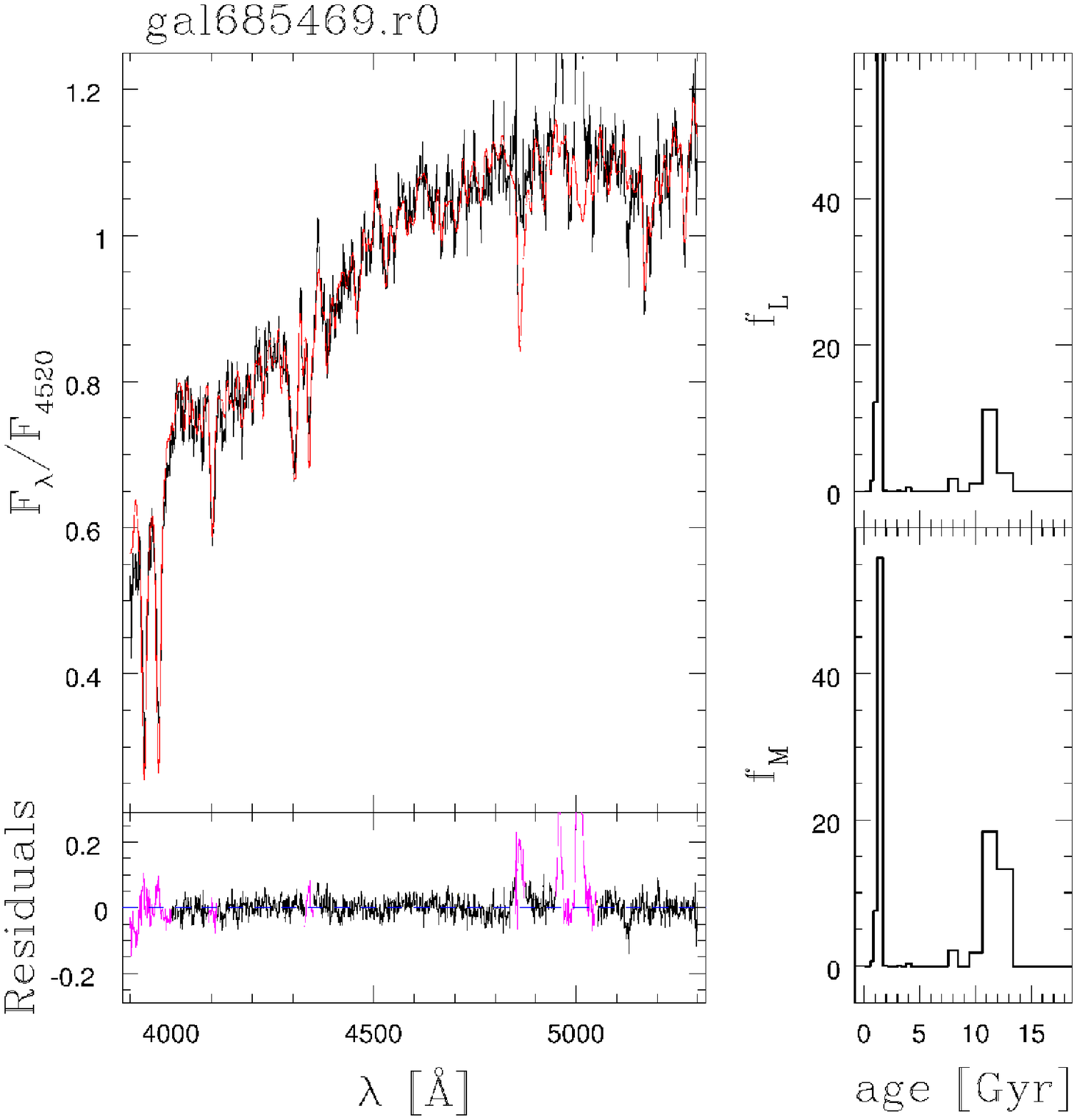} 
        \includegraphics[scale=0.28]{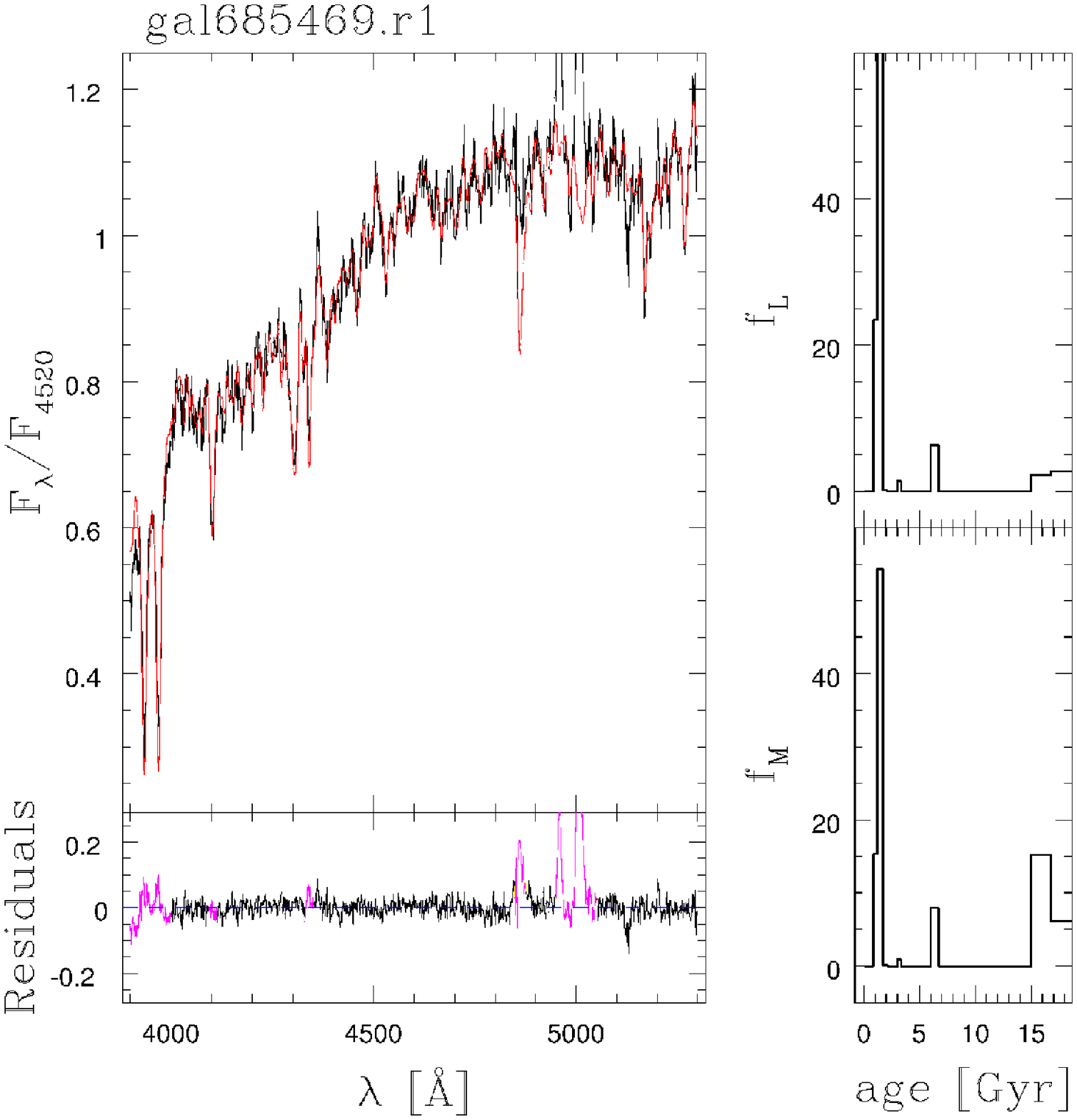} 
        \includegraphics[scale=0.28]{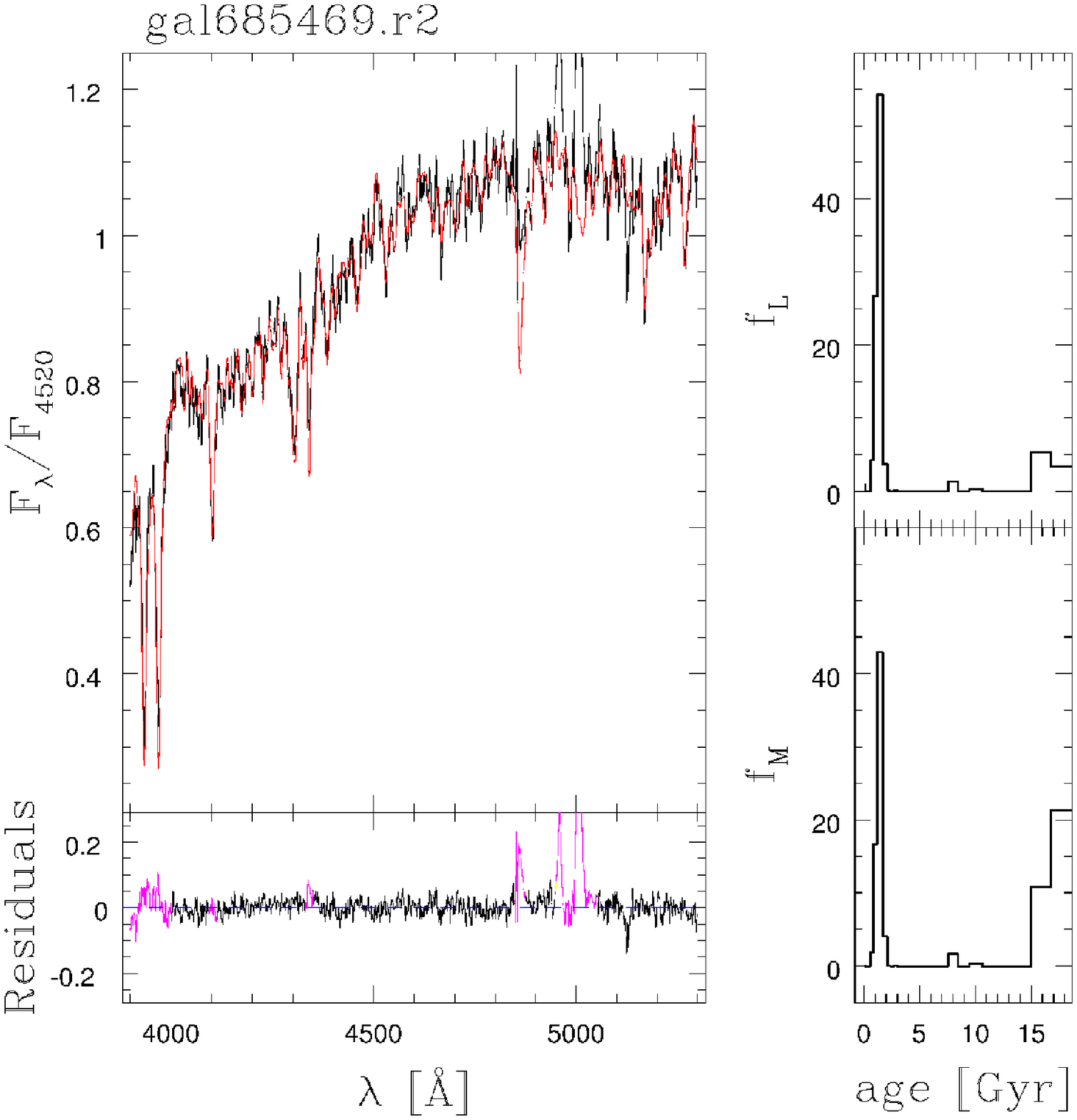} \\
        \includegraphics[scale=0.28]{gal79_sfh_r0_1.eps}
        \includegraphics[scale=0.28]{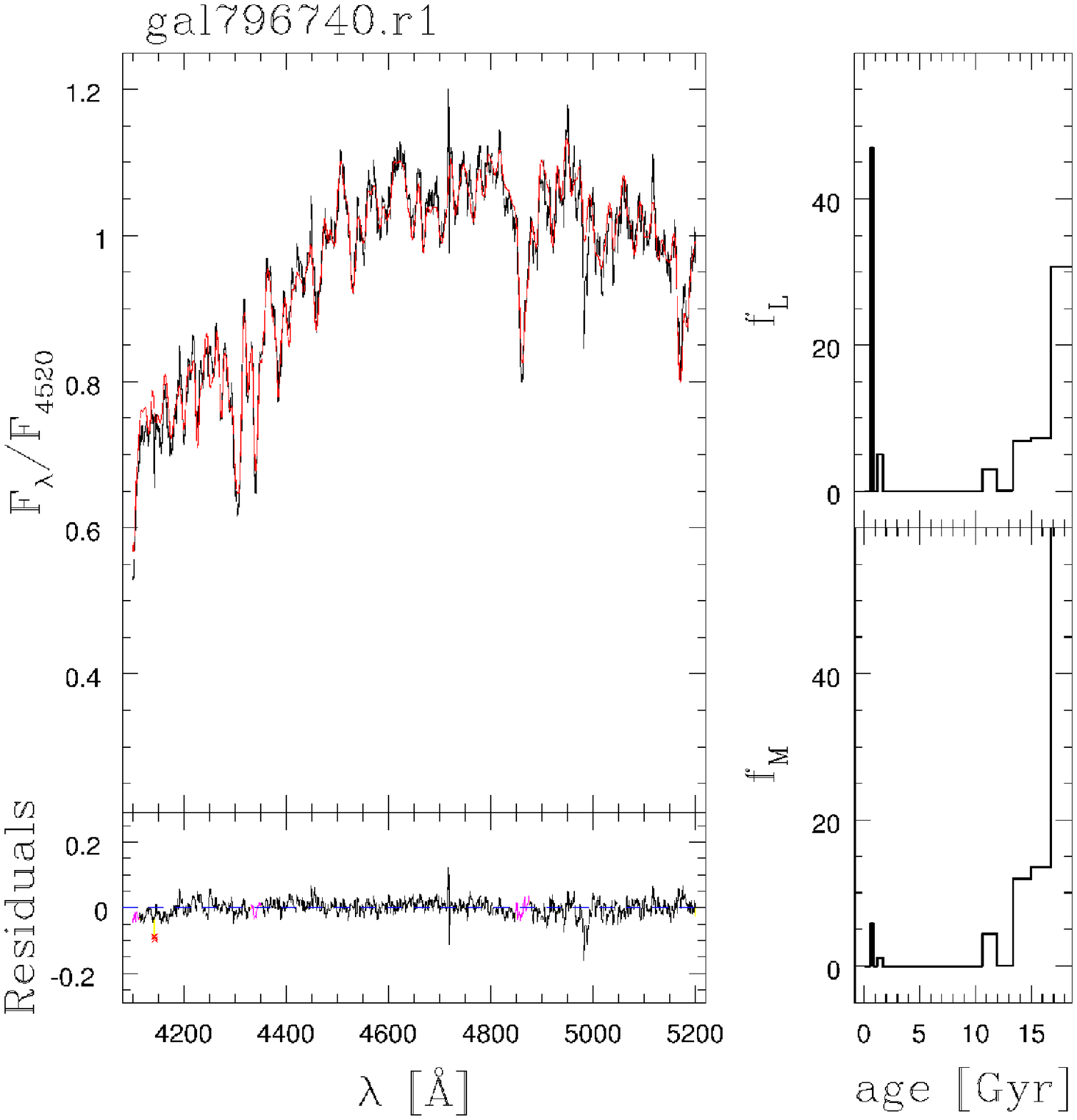}
        \includegraphics[scale=0.28]{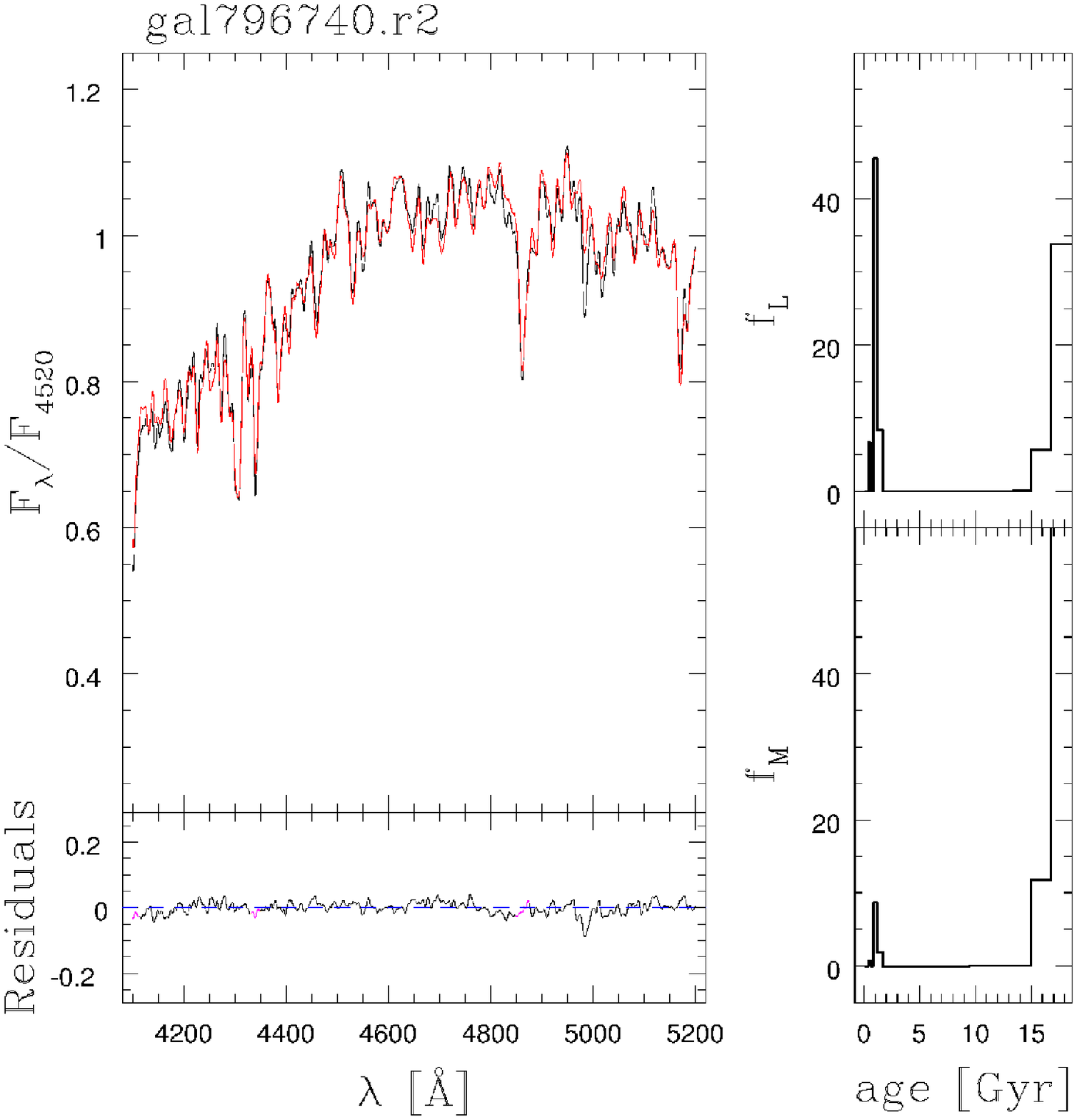} \\
\end{figure*} 

\begin{figure*}
  \begin{center}
        \includegraphics[scale=0.28]{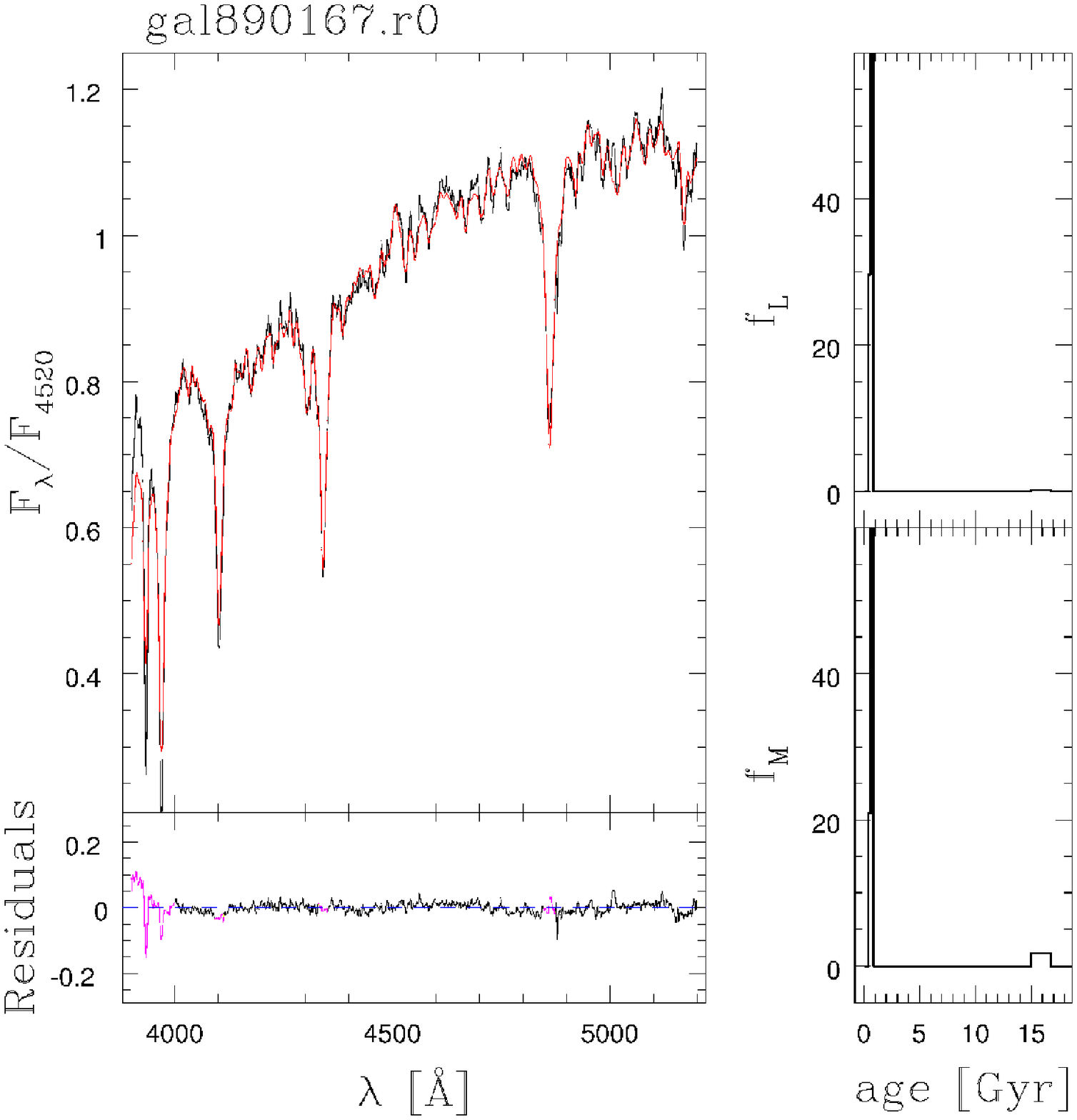}
        \includegraphics[scale=0.28]{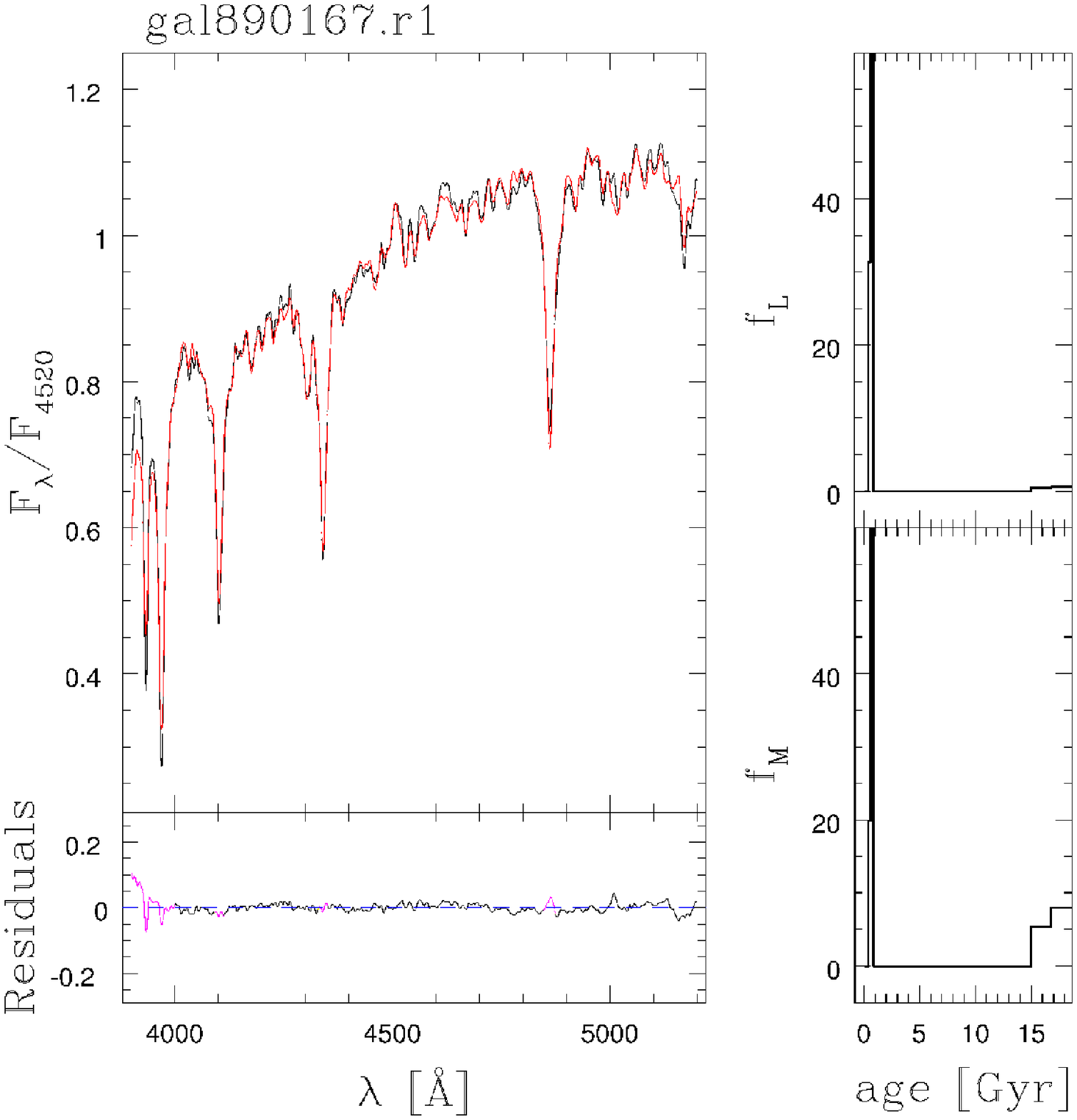}
        \includegraphics[scale=0.28]{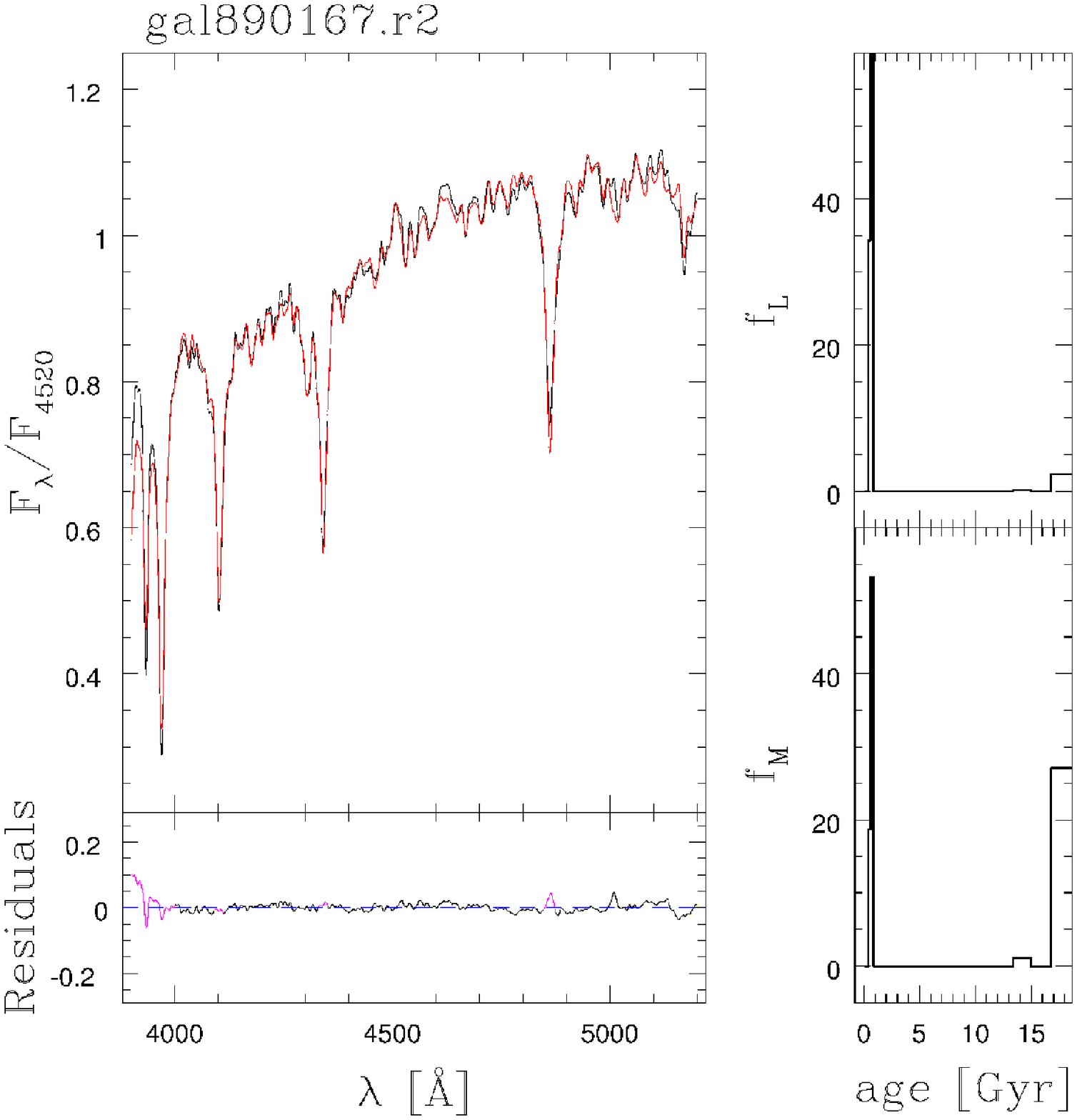} \\
        \includegraphics[scale=0.28]{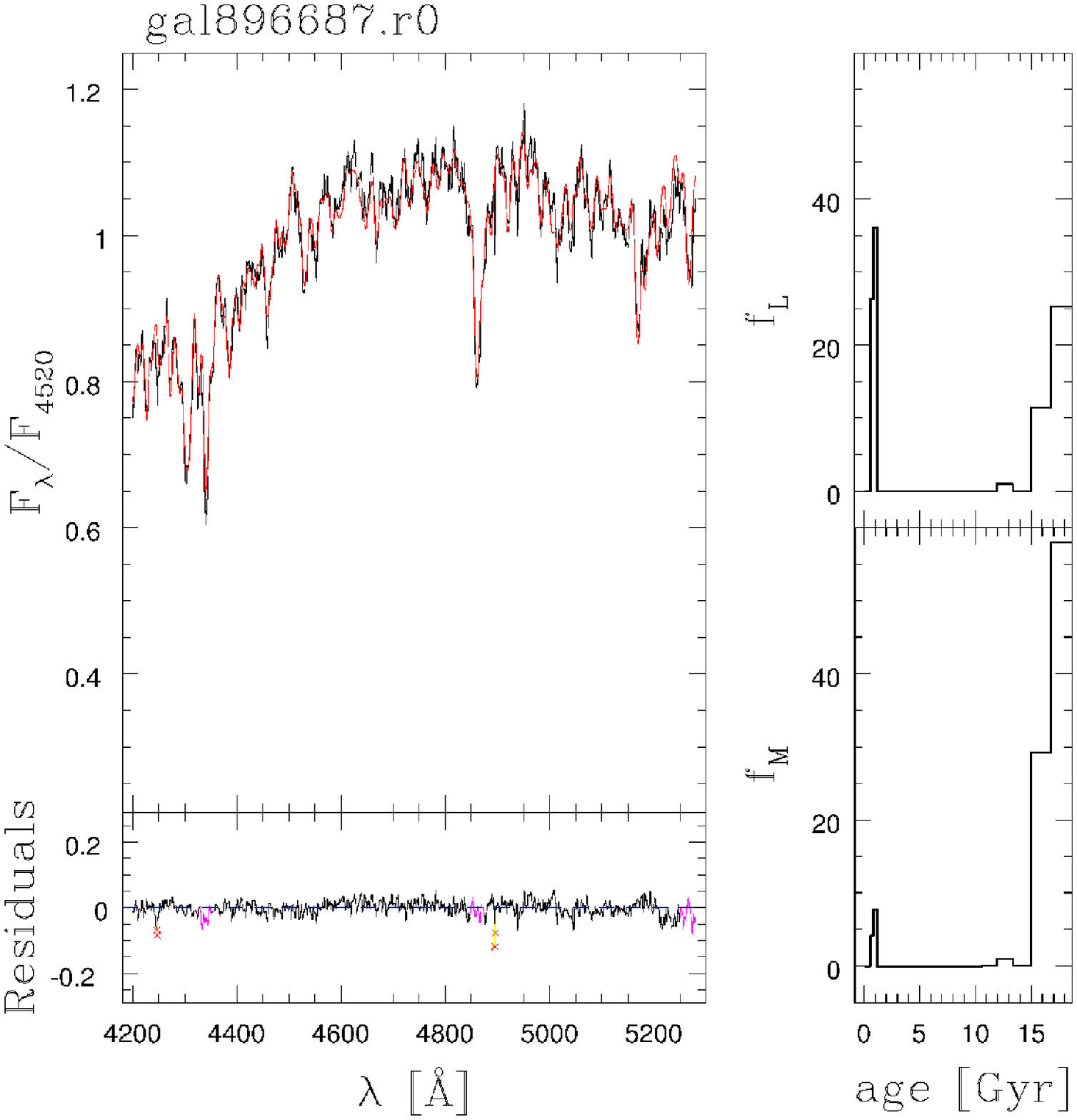}
        \includegraphics[scale=0.28]{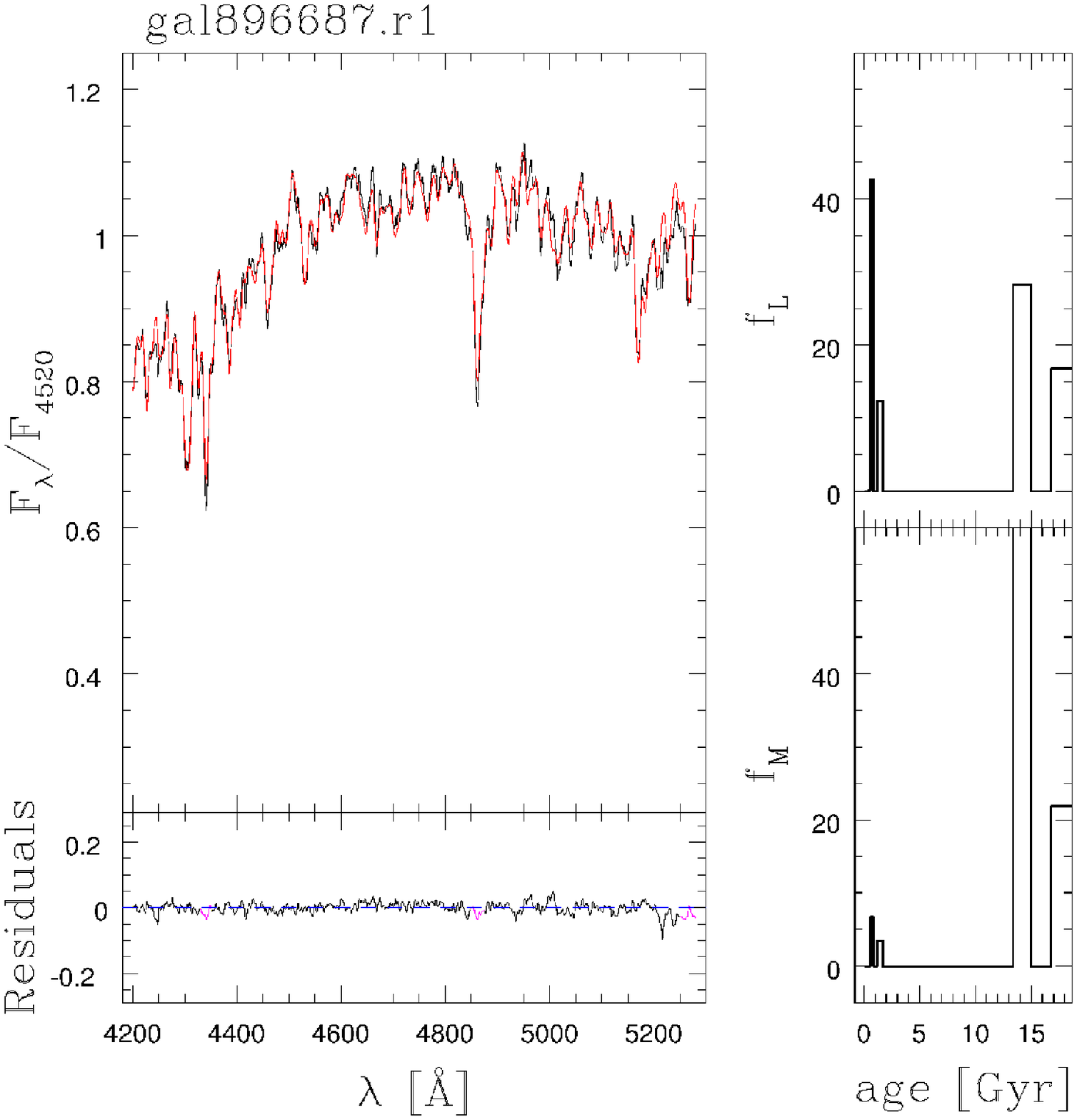}
        \includegraphics[scale=0.28]{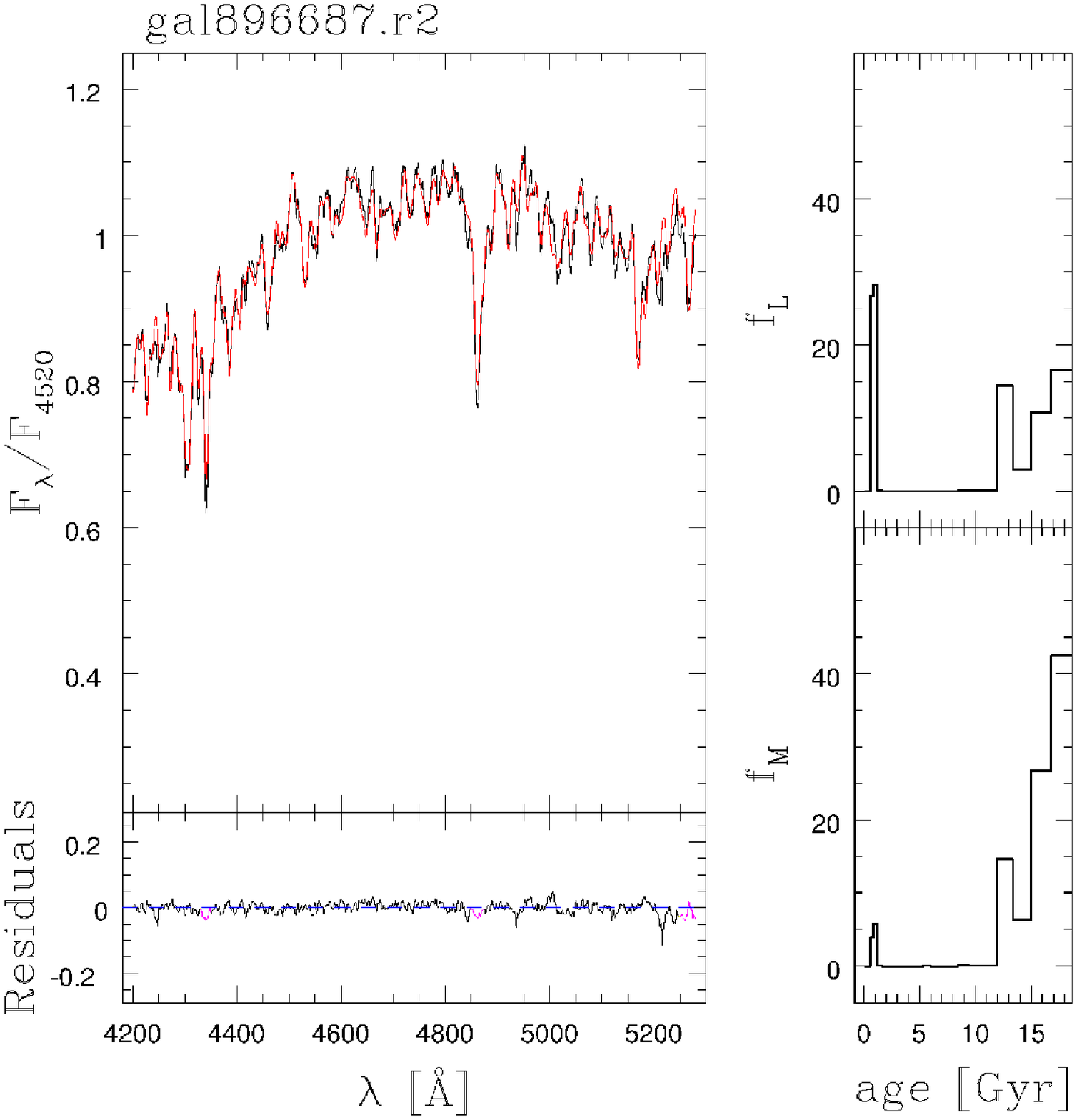} \\
        \includegraphics[scale=0.28]{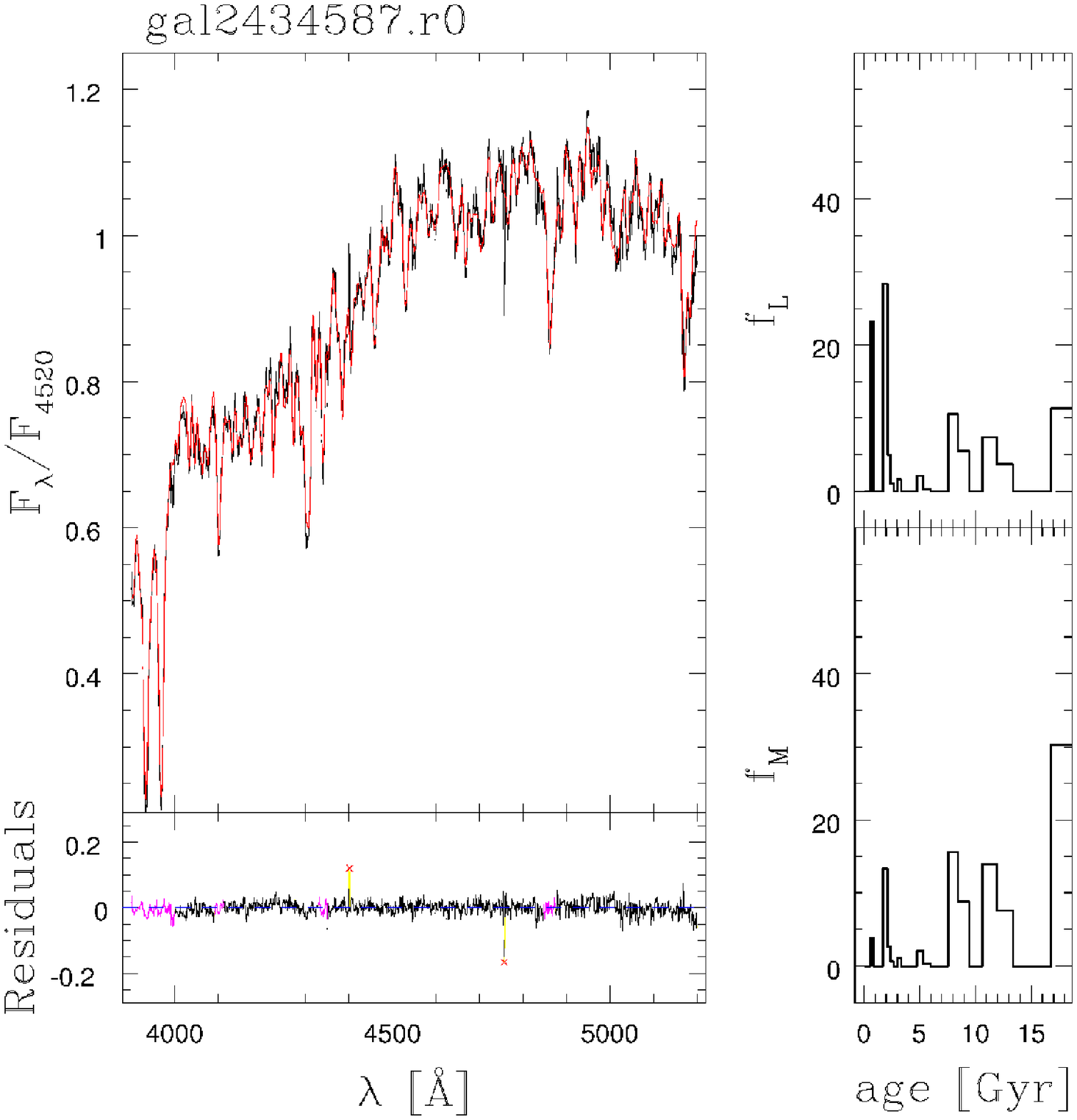}
        \includegraphics[scale=0.28]{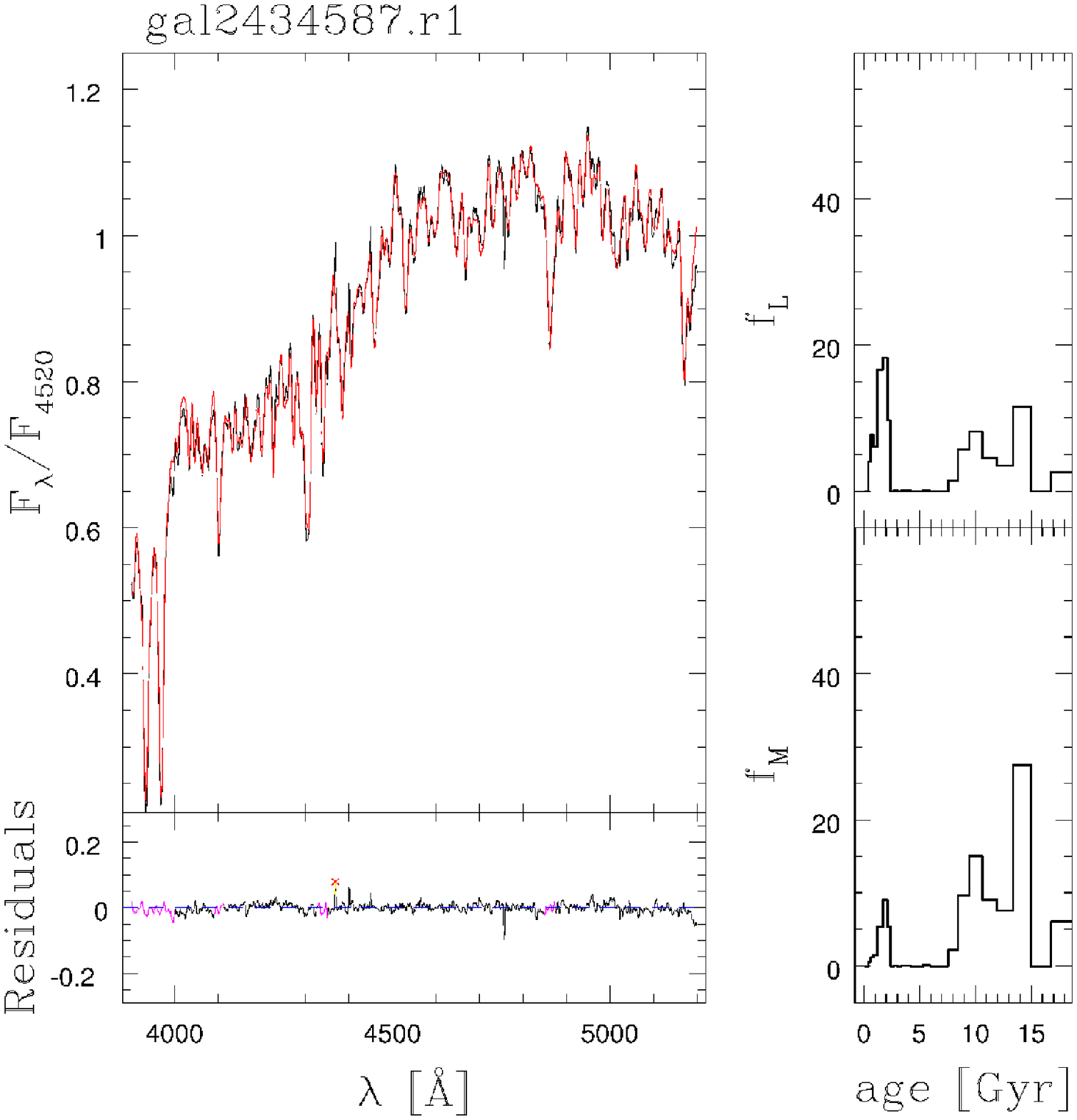}
        \includegraphics[scale=0.28]{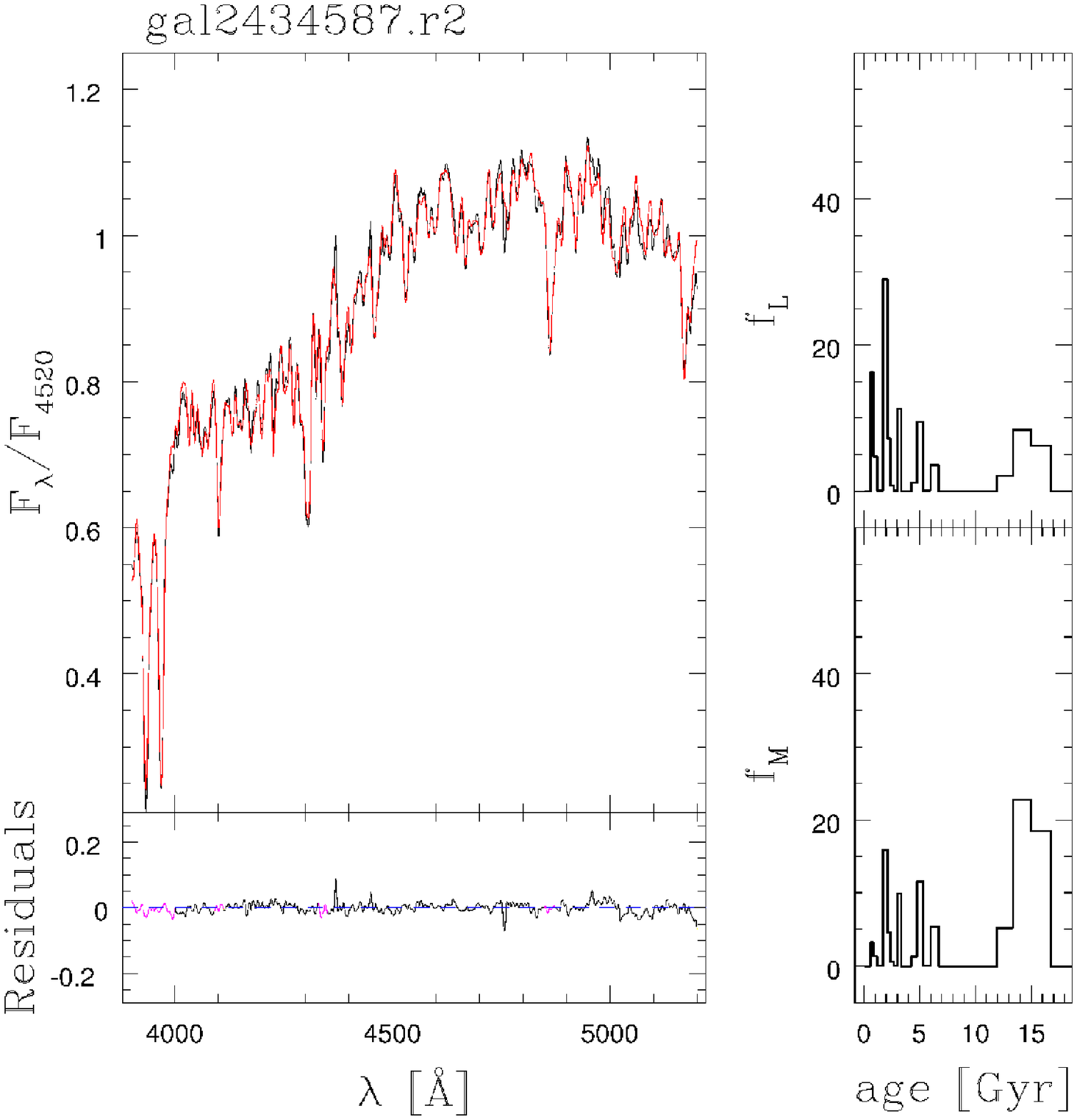} \\
\label{Fig.B1}
\caption{ Results from {\tt STARLIGHT} for the three apertures in each galaxy (r0, r1 and r2). On the upper-left hand, the galaxy spectra (black) is fitted with the models from V10 (red). The bottom-left panel shows the fitting residuals and the masked or rejected pixels (green). The right column on each panel shows the Star Formation History recovered in light (up) and in mass (down). It can be clearly seen that all our objects show recent starbursts, some of them leading to a big contribution of young populations.}
  \end{center}
 \end{figure*}

We have also tested the reliability of the full-spectral-fitting techniques to estimate ages and metallicities. We have applied two different tests, explained below.

TEST 1: This test constrains the fractions of young and old populations recovered with the full-spectral-fitting technique. Given the mean SSP-equivalent age and metallicity for a galaxy, derived from the SSPs (Single Stellar Population) model grids, we can also recover the same mean age with a combination of two SSPs, one younger and the other older than the SSP-age. These two SSPs can be weighted by different proportions, either in mass or in light. We created a sample of synthetic spectra with exactly the same configuration as for our galaxies, using the MILES webpage tool ``\textit{Get spectra from a given SFH}'', where our input were different SFHs, one containing a young population ($\lesssim$\,2\,Gyr) and the other containing an old population ($>$\,5\,Gyr), with varying proportions. We slightly vary the imposed limit (2, 1.5, 1 and 0.8\,Gyr) to match the estimated mean luminosity-weighted age for each object in our sample.
We then measure their line-strengths in the same way we did for our galaxies and plotted them in the H$\beta_{o}$ \textit{vs} [MgFe50] diagram (Figure B2a). The different combinations of ages with a fixed metallicity are plotted in different colours for [Z/H]=+0.22 and [Z/H]=0.0. Only those points (in black) that lie close to our galaxies are the possible SSP combinations that will give the galaxy mean age.
However, the dependence with the metallicity has also to be taken into account because of the known age-metallicity degeneracy (a young metal-rich system has a similar integrated energy distribution (SED) as an older metal-poor one). To illustrate this, we plot the extreme case where a young metal-rich population ([Z/H]=+0.22) is combined with an old metal-poor one ([Z/H]=-2.43). All the gray points correspond to those possible combinations of SSPs with different ages and different metallicities that lead to our galaxy values. As we know the amount of old population for each of them, we are able to impose limits on the fractions required to achieve our galaxies values.
At a fixed age, less metal-rich components contain smaller fraction of old populations. At fixed metallicity, as the mean luminosity-weighted age increases, the fraction of old population can reach any value. For example, for the galaxy 2434587 with mean SSP-age 2.24 Gyr, the amount of old population varies from 0.05$\%$ to 95$\%$, while for the galaxies with less than 1.0 Gyr, only SSPs with more than 60$\%$ of old population are feasible. The fractions listed in Table 4 are in agreement with this test, except for galaxy the 54829, whose young population is higher than expected.

\begin{figure*}
 \includegraphics[scale=0.6]{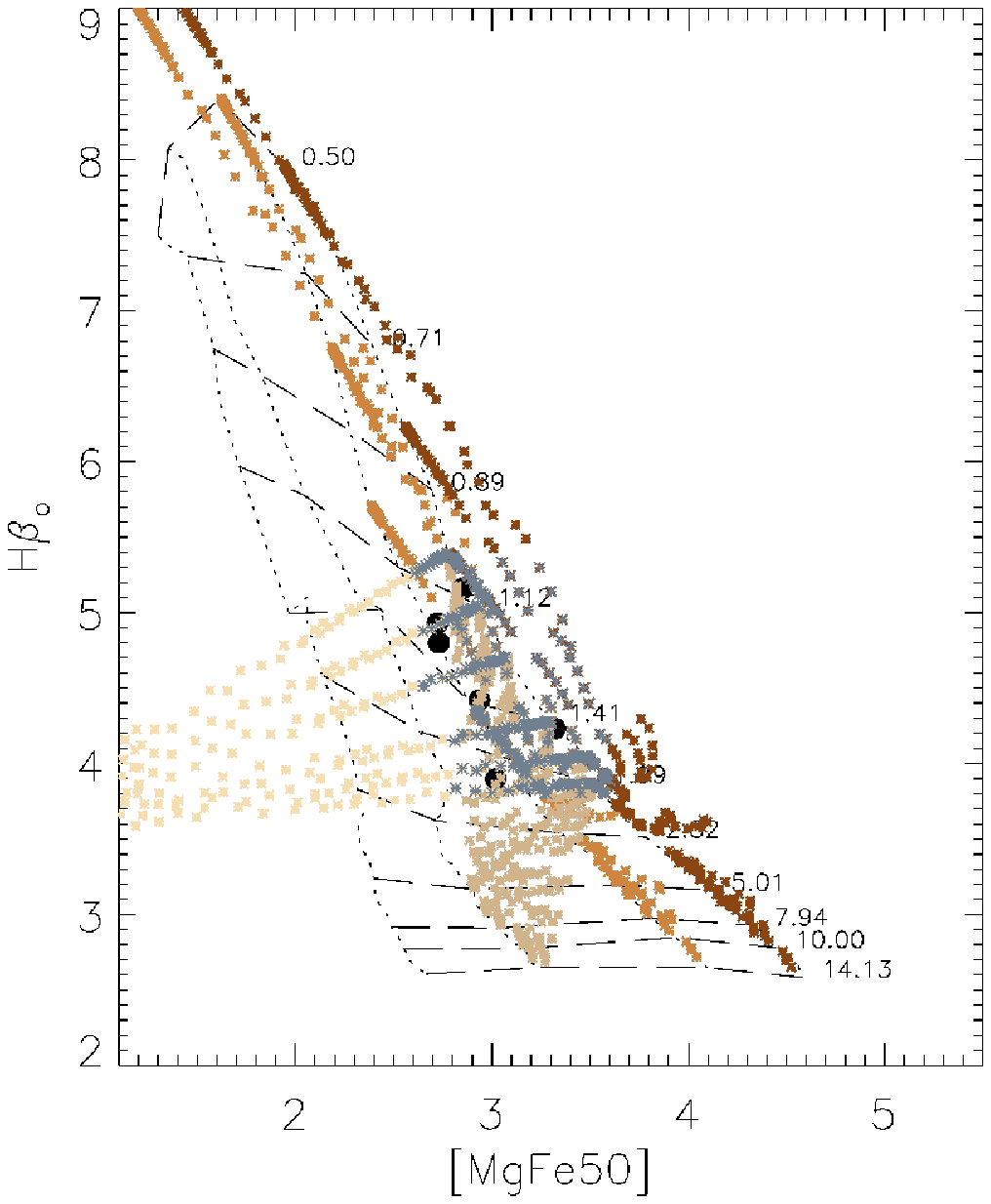}
 \includegraphics[scale=0.65]{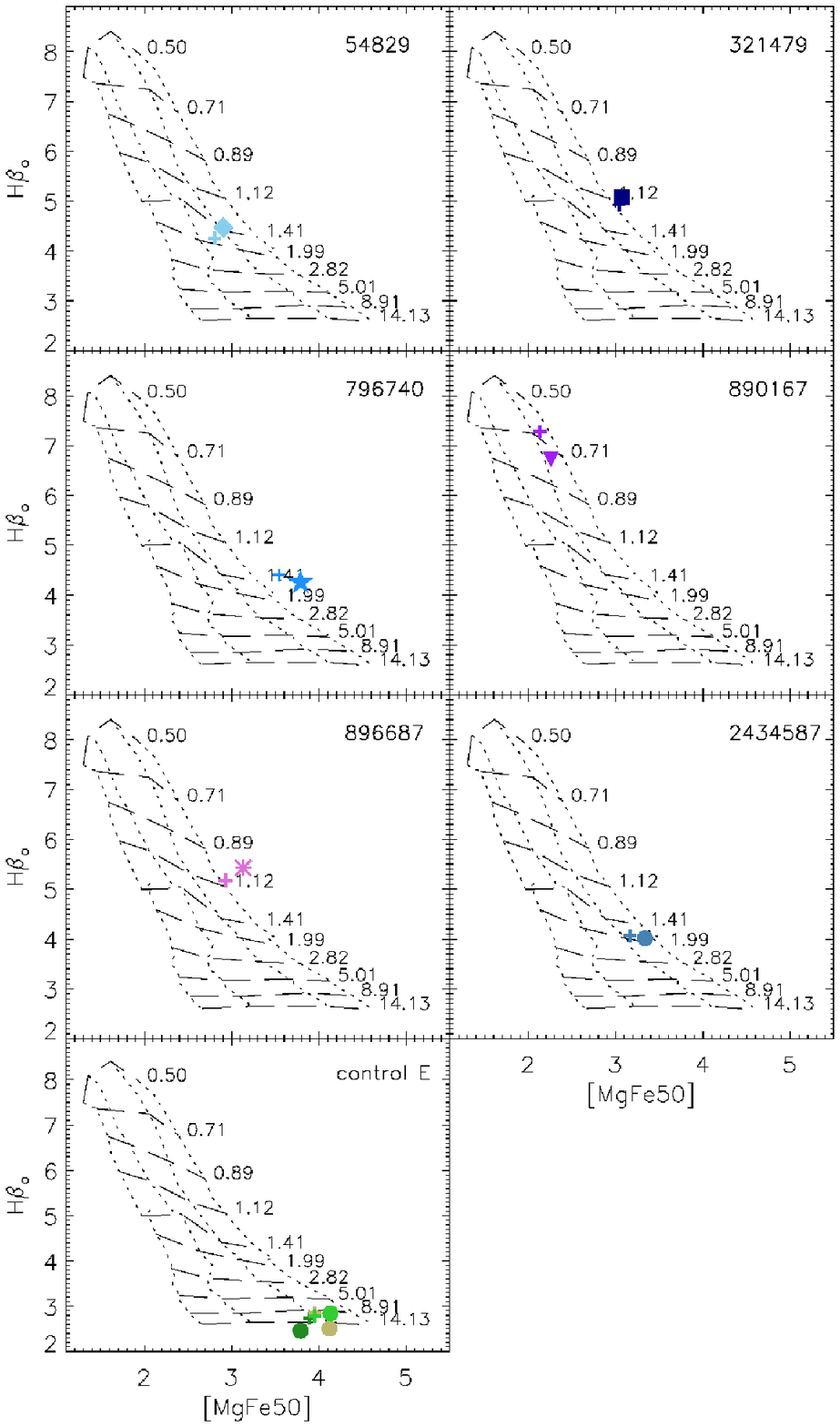}
 \caption{\textit{a. Left panel}: The age-sensitive indicator H$\beta{_o}$ is plotted against the composite metallicity index [MgFe50], to show the results of combining two different SSPs (young+old) with a fixed metallicity of [Z/H]= 0.0 and 0.22 (orange/brown symbols). We also show the effect of combining an old metal-poor with a young metal-rich SSP (horizontal sequence of faint points). In gray, we highlight the possible combinations that lead to the observed values. \textit{b. Right panel}: H$\beta{_o}$-[MgFe50] plots for the second test. Each panel refers to each galaxy of our sample. Large filled symbols are our compact galaxies as in Figure 2, while the crosses refer to the values derived from the fitted spectra from {\tt STARLIGHT}. For comparison we show the results for the control ellipticals in the bottom panel.}
\end{figure*}

TEST 2: This test is a simple reanalysis of the output of the full-spectral-fitting. We measured, in exactly the same way that we did for our galaxies, the absorption line-strengths of the best model fitted to each galaxy with {\tt STARLIGHT}. We plotted these values in the index-index diagrams to compare them with the observed value from the galaxy spectra (for both our galaxies and for the elliptical control sample). It appears (Fig. B2a) that ages are well recovered and metallicities tend to be slightly lower by -0.10Z$_{\sun}$. For old stellar populations, the method based on the Balmer indicator \textit{vs.} the metallicity indicator tend to provide older ages than the full-spectral-fitting approach (\citealt{Vazdekis2001}, \citealt{Mendel2007}). This effect is seen for the control elliptical galaxies on the bottom panel of the figure. Instead, in stellar populations with with strong contribution of young components, the SSP-equivalent age is strongly biased towards the age of these young components (e.g., \citealt{Serra2007}).\\

\label{lastpage}
\end{document}